\documentclass[a4paper,11pt]{article}
\pdfoutput=1 

\usepackage{jheppub} 

\usepackage[OT1]{fontenc} 
\graphicspath{{Figures/}} 	     
\usepackage{booktabs} 
\usepackage{bbold} 
\usepackage{mathtools}
\usepackage{multirow}
\usepackage[version=4]{mhchem} 
\usepackage{physics}
\usepackage{pdflscape}
\usepackage{slashed}
\usepackage{xcolor}
\usepackage{caption}
\usepackage{ragged2e}
\usepackage{orcidlink}
\usepackage{subcaption}

\newcommand{\bea} {\begin{eqnarray}}
\newcommand{\eea} {\end{eqnarray}}

\newcommand{\beq}{\begin{equation}}
\newcommand{\eeq}{\end{equation}}

\title{ Entanglement island and Page curve for one-sided rotating Kerr black holes }  

  \author[a,b]{Soumen Pari \orcidlink{0009-0009-1206-3109}}
\affiliation[a]{Theory division, Saha Institute of Nuclear Physics, 1/AF, Bidhannagar, West Bengal 700064, India.}
\affiliation[b]{Homi Bhabha National Institute, Training School Complex, Anushaktinagar, Mumbai 400094, India.}
    \emailAdd{soumen.pari@saha.ac.in} 
\abstract{ We investigate the information paradox for one-sided asymptotically flat rotating Kerr black holes formed through gravitational collapse using the island prescription. Near the horizon, the dynamics of a scalar field can be described by an effective two-dimensional theory, which provides the framework for our analysis. We study how rotation affects the location of the island boundary $\partial I$, the Page time, and the scrambling time, treating non-extremal and extremal Kerr black holes separately. For non-extremal black holes, the emergence of an island prevents the unbounded growth of Hawking radiation entropy at late times and restores the unitary Page curve. We further find that the island boundary depends on the location of the cutoff surface, in contrast to the behaviour found for eternal Kerr black holes. In the extremal case, the late-time entanglement entropy of Hawking radiation approaches the Bekenstein--Hawking entropy of the extremal black hole. Our results demonstrate the important role of black hole geometry in determining the behaviour of entanglement entropy and suggest that the island prescription provides a semiclassical resolution of the information paradox for one-sided rotating Kerr black holes.}

\begin{document}
\maketitle
\flushbottom 

\section{Introduction} \label{sec1}

\begin{figure}[!t]
    \centering
    \includegraphics[width=1.1\textwidth]{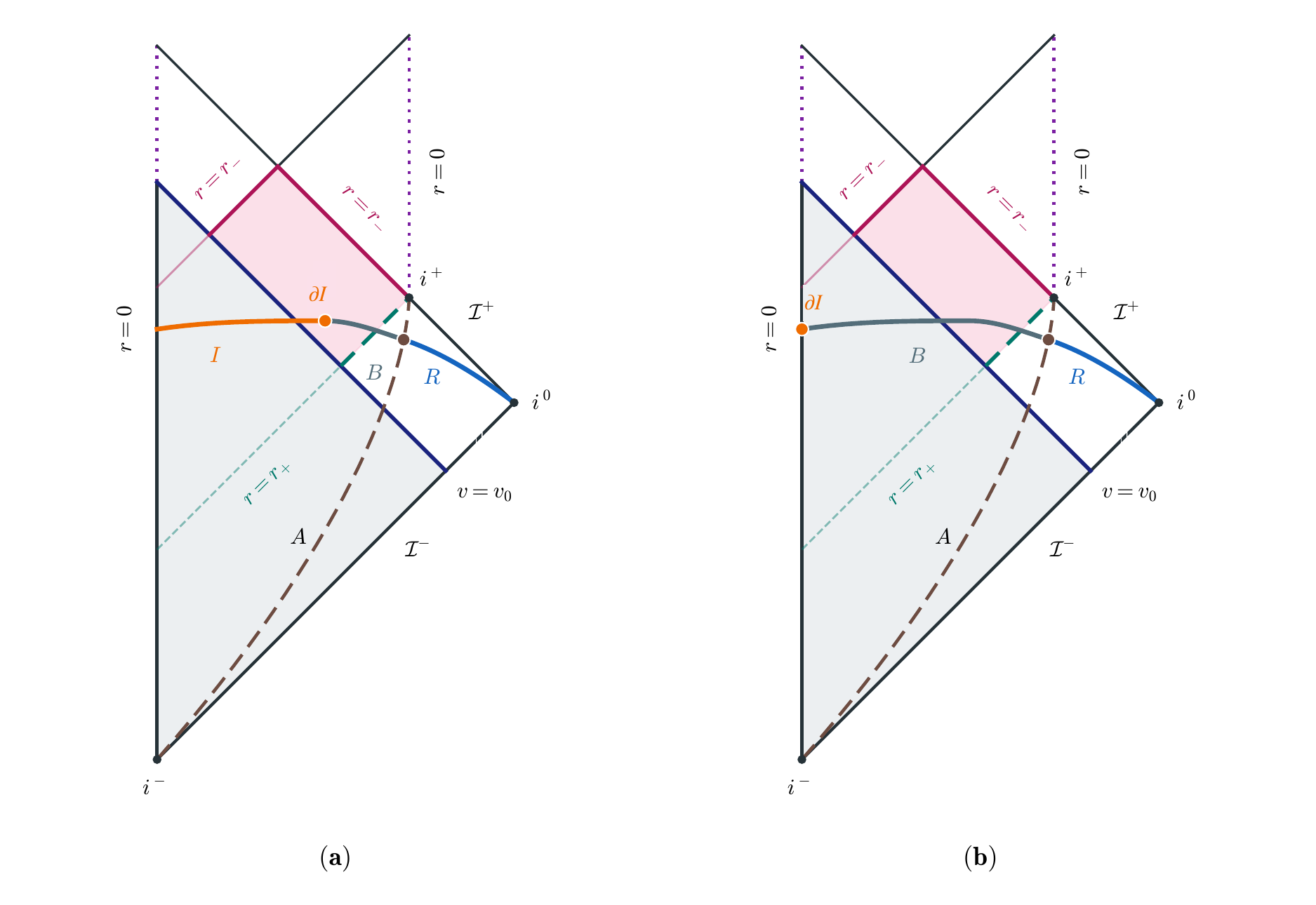}
  \caption{(a) Penrose diagram of a one-sided Kerr black hole formed by the collapse of a spherical null shell at $v=v_0$. The blue dashed and violet curves represent the event horizon, $r=r_+$, and the Cauchy horizon, $r=r_-$, respectively. An observer at the cutoff surface (brown dashed curve) collects Hawking radiation in the region $R$. The orange curve represents the island, with $\partial I$ denoting its boundary. (b) Configuration without an island, which can equivalently be described by setting $r_I=0$.}
    \label{fig:Penrose_diagram_non-extremal}
\end{figure}

The black hole information paradox \cite{Hawking:1976ra} has remained one of the longstanding challenges in semiclassical quantum gravity for more than four decades. It originates from Hawking's discovery that black holes emit thermal radiation due to quantum effects in curved spacetime \cite{Hawking:1975vcx}. This discovery established a connection between black holes and thermodynamics, allowing them to be described as physical systems with well-defined temperature and entropy \cite{Almheiri:2020cfm}. However, it also raises a fundamental question about the preservation of quantum information. Consider a black hole formed through the gravitational collapse of an initially pure quantum state \cite{Christodoulou:2008nj}. As the black hole evolves, it emits Hawking radiation, which is commonly understood as arising from entangled particle pairs created near the event horizon through vacuum fluctuations \cite{Hawking:1975vcx}. One member of each pair escapes towards infinity, while the other falls behind the horizon. Although the combined system remains in a pure state, an observer outside the black hole has access only to the outgoing radiation and therefore describes it as a mixed state. If the black hole evaporates completely, the final radiation appears to remain in a mixed state, despite the initially pure state. This apparent loss of information contradicts the principle of unitarity in quantum mechanics and lies at the heart of the black hole information paradox \cite{Hawking:1976ra}.\\
To investigate the information paradox, it is important to examine the entanglement entropy associated with a black hole. This entropy is described by the von Neumann entropy, also known as the fine-grained entropy. For a quantum system with density matrix $\rho$, it is defined as
\begin{equation}
S_{\mathrm{vN}}=-\mathrm{Tr}(\rho\ln\rho).
\end{equation}
For a pure state, the von Neumann entropy vanishes and remains invariant under unitary time evolution. Thus, if a black hole forms from the gravitational collapse of an initially pure state, the fine-grained entropy of the complete system should remain zero throughout its evolution. However, Hawking's semiclassical analysis predicts that the entanglement entropy of the emitted radiation increases linearly with time during black hole evaporation \cite{Hawking:1976ra}. At sufficiently late times, this entropy can exceed the Bekenstein--Hawking entropy of the black hole. Such behaviour raises a fundamental concern, since the thermodynamic entropy is expected to account for the number of available microscopic degrees of freedom and hence constrain the fine-grained entropy of the corresponding system.\\
\begin{figure}[t]
    \centering
    \includegraphics[width=0.65\textwidth]{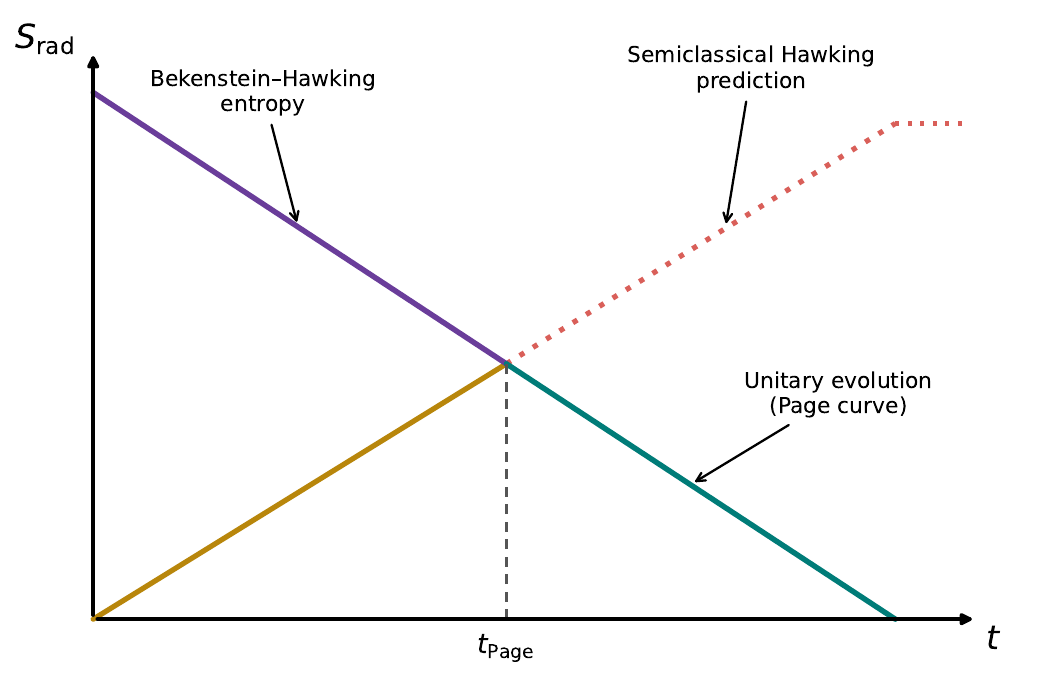}
    \caption{Schematic illustration of the time evolution of the
    entanglement entropy of Hawking radiation. The semiclassical
    Hawking prediction exhibits a continuous increase, whereas
    the unitary evolution leads to the Page curve, in which the
    entropy decreases after reaching its maximum at the Page time.}
    \label{fig:Page_curve}
\end{figure}
Nevertheless, black holes are generally expected to obey the principle of unitarity when treated as quantum systems. Page investigated this possibility and proposed a characteristic time evolution of the radiation entropy, now known as the Page curve \cite{Page:1993wv,Page:2013dx}. According to this picture, the entanglement entropy of Hawking radiation initially increases with time, reaches a maximum at the Page time, and subsequently decreases, eventually approaching zero after complete evaporation. This behaviour differs from Hawking's semiclassical prediction, which suggests a continuous increase in radiation entropy. The two predictions are illustrated schematically in Fig.~\ref{fig:Page_curve}. Reproducing the Page curve within a gravitational framework has therefore become an important step towards understanding the black hole information paradox \cite{Hawking:1976ra}. Such a result would provide strong evidence for the consistency of unitary evolution in the corresponding gravitational description.\\
The assumption of unitary evolution during black hole evaporation gained strong theoretical support with the advent of the AdS/CFT correspondence \cite{Maldacena:1997re}. This duality relates quantum gravity in an asymptotically AdS spacetime to a conformal field theory (CFT) defined on its boundary. Since the boundary CFT evolves unitarily, the corresponding bulk gravitational theory is also expected to preserve quantum information. In particular, the evaporation of black holes in an AdS spacetime can be understood through the unitary dynamics of the dual boundary theory. Although this viewpoint has been challenged by the firewall paradox \cite{Almheiri:2012rt}, the AdS/CFT correspondence provides a powerful framework for understanding the preservation of information and addressing the black hole information paradox.\\
A major development in addressing the black hole information paradox came from the calculation of Hawking radiation entropy and the reproduction of the Page curve within the framework of the AdS/CFT correspondence \cite{Penington:2019npb,Almheiri:2019psf,Almheiri:2019hni,Penington:2019kki,Almheiri:2019qdq}. This progress builds upon the pioneering Ryu--Takayanagi (RT) prescription, which relates the entanglement entropy of a boundary region to the area of a minimal surface in the bulk spacetime \cite{Ryu:2006bv}. Incorporating quantum corrections led to the development of the quantum extremal surface (QES) prescription \cite{Hubeny:2007xt,Lewkowycz:2013nqa,Barrella:2013wja,Engelhardt:2014gca,Faulkner:2013ana}. An important consequence of this development is the island rule, which provides a prescription for computing the fine-grained entropy of Hawking radiation by including contributions from additional bulk regions known as islands. According to the island prescription, the entropy of a black hole and that of its associated radiation can be expressed as
\begin{equation}\label{eq:sgen for BH}
S(BH)=\underset{I}{\mathrm{min}}
\left\{
\underset{\partial I}{\mathrm{ext}}
\left[
\frac{\mathrm{Area}(\partial I)}{4G_N}
+S_{\mathrm{ent}}(B)
\right]
\right\},
\end{equation}
and
\begin{equation}\label{eq:sgen for R}
S(R)=\underset{I}{\mathrm{min}}
\left\{
\underset{\partial I}{\mathrm{ext}}
\left[
\frac{\mathrm{Area}(\partial I)}{4G_N}
+S_{\mathrm{ent}}(I\cup R)
\right]
\right\},
\end{equation}
respectively. Here, $R$ denotes the radiation region outside the cutoff surface $A$, while $I$ represents a bulk island whose boundary is denoted by $\partial I$. The region $B$ is bounded by the island boundary $\partial I$, and the cutoff surface $A$ (Fig.~\ref{fig:Penrose_diagram_non-extremal}). The surface $\partial I$ is identified with the QES, whose location is determined by extremizing the generalized entropy \cite{Engelhardt:2014gca,Lewkowycz:2013nqa,Faulkner:2013ana,Dong:2016hjy}, defined as
\begin{equation}
S_{\mathrm{gen}}=
\frac{\mathrm{Area}(\partial I)}{4G_N}
+S_{\mathrm{ent}}.
\end{equation}
The first contribution represents the geometric entropy associated with the island boundary, while the second accounts for the entanglement entropy of the quantum matter fields. One first determines the possible extremal configurations of $S_{\mathrm{gen}}$ and then selects the one with the smallest generalized entropy. This prescription determines the fine-grained entropy of the Hawking radiation.

Furthermore, the island formula can be derived using the Euclidean gravitational path integral on a replicated manifold, without relying explicitly on the AdS/CFT correspondence \cite{Almheiri:2019qdq,Penington:2019kki}. In this approach, replica wormholes emerge as relevant saddle-point geometries of the gravitational path integral and provide a geometric explanation for the appearance of islands. This derivation has also been extended beyond eternal black holes to include evaporating black-hole geometries \cite{Goto:2020wnk,Hartman:2020swn}. More recently, the RT prescription in AdS/BCFT, for an asymptotically AdS$_3$ bulk geometry, has been derived using the replica path integral. In this setting, the topological transition of the RT surface has been associated with the formation of replica wormholes in the Karch-Randall braneworld construction \cite{Geng:2024xpj}.\\
The island prescription has been studied in a wide range of gravitational 
settings. In asymptotically AdS spacetimes, these include two-dimensional 
Jackiw-Teitelboim (JT) gravity coupled to a thermal bath 
\cite{Chen:2019uhq,Almheiri:2019hni,Almheiri:2020cfm}, de Sitter (dS) 
spacetime \cite{Geng:2021wcq,Balasubramanian:2020xqf}, curvature-squared 
gravity \cite{Hu:2022ymx}, charged black holes 
\cite{Ling:2020laa,Kim:2021gzd}, and higher-dimensional theories 
\cite{Almheiri:2019psy}. The prescription has also been extended to 
asymptotically flat spacetimes, including dilaton gravity 
\cite{Yu:2021cgi,Anegawa:2020ezn,Hartman:2020swn,Gautason:2020tmk}, 
flat-space cosmology \cite{Azarnia:2021uch}, exactly solvable 
two-dimensional black holes \cite{Wang:2021mqq}, Schwarzschild black 
holes \cite{Hashimoto:2020cas,Arefeva:2021kfx,Matsuo:2020ypv,Dong:2020uxp}, 
Reissner--Nordström (RN) black holes 
\cite{Wang:2021woy,Kim:2021gzd}, and eternal black holes 
\cite{He:2021mst}. This list is by no means exhaustive, as the 
island prescription continues to attract considerable attention 
across different gravitational settings.\\
Despite the extensive progress in this direction, relatively little 
attention has been devoted to rotating black holes, primarily due to 
the additional complexities associated with their spacetime geometry. 
The Kerr solution describes a rotating black hole characterized by 
its mass and angular momentum. Its distinctive features include a 
ring singularity and an ergoregion surrounding the event horizon. 
Since rotating black holes are expected to be of considerable 
relevance in realistic four-dimensional astrophysical settings, 
understanding the information paradox in the Kerr background is 
an important problem.\\
To the best of our knowledge, existing studies of rotating Kerr black holes \cite{Wang:2024itz,Yu:2021rfg,Das:2026cjf} and Kerr-Newman black holes \cite{Yu:2025euq} have primarily focused on eternal black-hole geometries and static vacuum states \cite{Matsuo:2021mmi}. In this context, the Hartle-Hawking state is particularly relevant\footnote{The Hartle–Hawking state is the unique quantum state that remains regular throughout the maximally extended Kruskal spacetime and is invariant under Schwarzschild time translations.}. Physically, it describes a black hole in thermal equilibrium with its surrounding radiation: outgoing Hawking radiation reaches 
future null infinity, $\mathcal{I}^{+}$, while an equal incoming 
thermal flux arrives from past null infinity, $\mathcal{I}^{-}$.\\
In this work, we consider the ``in'' vacuum state to describe a 
one-sided, dynamical Kerr black hole formed through the collapse 
of a spherical null shell (see Fig.~\ref{fig:Penrose_diagram_non-extremal}). Related studies of one-sided asymptotically flat black holes and charged 
Reissner-Nordström black holes formed in the ``in'' vacuum state 
can be found in \cite{Gan:2022jay,Qu:2024sby}. The ``in'' vacuum 
state is particularly suitable for describing black holes formed 
by gravitational collapse, as it is characterized by the absence 
of incoming radiation from past null infinity, $\mathcal{I}^{-}$. 
Unlike the Hartle-Hawking state, which describes a black hole in 
thermal equilibrium with its surrounding radiation, the ``in'' 
vacuum state provides a more appropriate description of a 
one-sided, dynamically formed Kerr black hole. This choice of 
quantum state is a key distinction between our analysis and 
previous studies of rotating Kerr black holes 
\cite{Wang:2024itz,Yu:2021rfg,Das:2026cjf}, which considered 
the Hartle-Hawking state.\\
In this work, we investigate the island configurations and Page curve 
for a rotating Kerr black hole formed through the gravitational 
collapse of a spherical null shell, using the island prescription 
in the ``in'' vacuum state. Since obtaining analytical expressions 
for entanglement entropy in higher-dimensional spacetimes is 
generally difficult, the s-wave approximation is commonly employed 
to reduce the problem to an effective two-dimensional matter theory. 
In the case of four-dimensional Kerr black holes, the scalar field 
admits an effective two-dimensional description in the near-horizon 
limit \cite{Yu:2025euq,Umetsu:2009ra}. Accordingly, we formulate 
our analysis within this reduced two-dimensional framework, where 
the s-wave approximation provides an appropriate description of 
the matter sector. Throughout this work, we assume $cG \ll r_+^2$, as appropriate for a large macroscopic black hole, and neglect the backreaction of Hawking radiation on the background geometry.\\
The rotation of a Kerr black hole can trigger superradiance, potentially leading to spacetime instabilities \cite{Braidotti_2024,Teukolsky:1974yv,Unruh:1974bw}. To simplify our analysis, we consider the regime in which the black hole mass is much larger than its angular momentum, so that Hawking radiation dominates over superradiant effects. Our main objective is to examine how the rotation of the black hole 
affects the location of the entanglement island, the Page time, and the scrambling time. We further extend our analysis to extremal Kerr black holes, whose geometric and causal properties differ significantly from those of their non-extremal counterparts. By studying these two cases, we aim to clarify the role of rotation and extremality in the evolution of radiation entropy and the 
associated information-theoretic properties of one-sided Kerr black holes.\\
The remainder of this paper is organized as follows. In Section~\ref{sec2}, we review the basic properties of the Kerr black hole and introduce the dimensional reduction procedure that provides an effective two-dimensional description of the four-dimensional Kerr geometry near the event horizon. This 
description serves as the basis for our subsequent analysis of entanglement entropy. In Section~\ref{sec3}, we present the theoretical framework for calculating entanglement entropy and discuss its dependence on the choice of quantum state. In Section~\ref{sec 4}, we investigate the entanglement entropy for non-extremal Kerr black holes and examine the effects of angular 
momentum on the Page time and scrambling time. We consider two distinct cases depending on the location of the cutoff surface. When it is far from the horizon, the island boundary $\partial I$ lies just inside the horizon. In contrast, when the cutoff surface is close to the horizon, $\partial I$ is located just outside it. In Section~\ref{sec 6}, we extend our analysis to extremal Kerr black holes formed through gravitational collapse and find that their behaviour differs from that of non-extremal Kerr black holes. Nevertheless, in both cases, the island prescription ensures that the radiation entropy remains bounded. In Section~\ref{sec compareison}, we compare our results with previous studies of eternal Kerr black holes. Finally, in Section~\ref{sec 7}, we summarize our main results and discuss possible directions for future research.

\section{Kerr black hole and its effective description}\label{sec2}
\subsection{Kerr black hole}\label{subsec1}

The four-dimensional Kerr black-hole solution in Boyer–Lindquist coordinates is given by \cite{Wang:2024itz}
\begin{equation}
\label{eq:Kerr_metric}
\begin{aligned}
ds^2 =\; &- \frac{\Delta - a^2 \sin^2 \theta}{\Sigma} \, dt^2
- \frac{2a \sin^2 \theta (r^2 + a^2 - \Delta)}{\Sigma} \, dt \, d\phi \\
&+ \frac{(r^2 + a^2)^2 - \Delta a^2 \sin^2 \theta}{\Sigma} \, \sin^2 \theta \, d\phi^2
+ \frac{\Sigma}{\Delta} \, dr^2
+ \Sigma \, d\theta^2,
\end{aligned}
\end{equation}
where
\begin{equation}
\label{eq:parameters}
\Sigma = r^2 + a^2 \cos^2 \theta,\qquad
\Delta = r^2 - 2Mr + a^2 = (r - r_+)(r - r_-),\qquad
a=\frac{J}{M}.
\end{equation}
Here, \(M\) and \(J\) denote the mass and angular momentum of the black hole, respectively, while \(a\) is the rotation parameter. The locations of the two horizons are determined by the roots of \(\Delta=0\) : 
\begin{equation}
\label{roots}
r_{\pm} = M\pm\sqrt{M^2-a^2}.
\end{equation}
For \(M>a>0\), the two roots are distinct, corresponding to the outer event horizon \(r_+\) and the inner Cauchy horizon \(r_-\). The black hole is then non-extremal. In the limiting case \(M=a\), the two horizons coincide, and the solution becomes extremal Kerr black holes.

The surface gravity associated with the outer horizon is
\begin{equation}
\label{Surface gravity}
\kappa(r) = \sqrt{-\frac{1}{2}\nabla^{\mu}\zeta^{\nu}\nabla_{\mu}\zeta_{\nu}}=  \frac{r_+ - r_-}{2 (r_+^2 + a^2)}.
\end{equation}
The corresponding horizon-generating Killing vector is
\(\zeta^{\mu}=(\partial_{t})^{\mu}+\Omega_{H}(\partial_{\phi})^{\mu}\), where
\(\Omega_{H}=\frac{a}{r_{+}^2+a^2}\) is the angular velocity of the event horizon. Here, \(\nabla_{\mu}\) denotes the covariant derivative. The Hawking temperature is consequently given by
\begin{equation}
\label{eq:Hawking temp}
T_H = \frac{\kappa(r)}{2\pi}= \frac{r_+ - r_-}{4\pi(r_+^2 + a^2)}.
\end{equation}

The area of the outer event horizon follows from integrating the induced volume element over the angular coordinates:
\begin{equation}
\mathcal{A}=\int\sqrt{g_{\theta\theta}g_{\phi\phi}} \ d\theta d\phi|_{r=r_+}=\int_{0}^{2\pi}d\phi\int_{0}^{\pi}d\theta(r_{+}^2+a^2)\sin\theta=4\pi(r_{+}^2+a^2).
\end{equation}
The associated Bekenstein–Hawking entropy is therefore
\begin{equation}\label{eq:BH entropy}
S_{BH} = \frac{\mathcal{A}}{4G_{N}}=\frac{\pi (r_+^2 + a^2)}{G_N}.
\end{equation}
These quantities describe the basic thermodynamic properties of the Kerr black hole. In particular, the extremal limit \(r_+=r_-\) gives \(\kappa=0\), and hence a vanishing Hawking temperature \eqref{eq:Hawking temp}, while the horizon area and Bekenstein–Hawking entropy \eqref{eq:BH entropy} remain finite.
 
\subsection{Two-dimensional reduction of  Kerr geometry}\label{subsec2}

To investigate the near-horizon physics of a four-dimensional Kerr black hole, we consider an effective description in two dimensions. Following the approach of \cite{Yu:2025euq}, we show how the near-horizon limit allows the four-dimensional quantum field theory to be reduced to an effective two-dimensional theory. This reduction provides the two-dimensional metric and the corresponding effective description needed for our subsequent analysis.\\
The action for a massless complex scalar field $\varphi$ in the
four-dimensional Kerr spacetime is given by
\begin{equation}
\label{eq:scalar action}
S[\varphi]=\int_{V} d^4x\sqrt{-g}\,
g^{\mu\nu}\partial_{\mu}\varphi^{*}\partial_{\nu}\varphi,
\end{equation}
where $g$ denotes the determinant of the four-dimensional Kerr
metric. Substituting the metric \eqref{eq:Kerr_metric} into the
action \eqref{eq:scalar action} and performing the necessary
integrations by parts, we obtain \footnote{This action agrees with the complex scalar field action for Kerr Newman spacetime derived in \cite{Yu:2025euq} in the uncharged limit, $Q=0$.}
\begin{align}
S[\varphi] =\;& \int dtdrd\theta d\phi\,\sin\theta\,\varphi^{*}
\Bigg\{
\left(
\frac{(r^2+a^2)^2}{\Delta}
-a^2\sin^2\theta
\right)\partial_t^2
\nonumber\\
&\quad +2ia\left(
\frac{r^2+a^2}{\Delta}-1
\right)\partial_t\hat{L}_z
+\hat{L}^2
-\partial_r\left(\Delta\partial_r\right)
-\frac{a^2}{\Delta}\hat{L}_z^2
\Bigg\}\varphi,
\end{align}
where we have introduced the angular momentum operators
\begin{equation}
\hat{L}=-\frac{1}{\sin\theta}\partial_{\theta}
\sin\theta\partial_{\theta},
\qquad
\hat{L}_z=-i\partial_{\phi}.
\end{equation}

Using the spherical harmonics expansion
$\varphi=\sum_{\ell,m}\varphi_{\ell m}(t,r)Y_{\ell m}(\theta,\phi)$,
we obtain
\begin{align}\label{eq:action after}
S[\varphi] =\;& \int dtdrd\theta d\phi\,\sin\theta\,
\sum_{\ell',m'}\varphi_{\ell' m'}^{*}Y_{\ell' m'}^{*}
\Bigg\{
\left(
\frac{(r^2+a^2)^2}{\Delta}
-a^2\sin^2\theta
\right)\partial_t^2
\nonumber\\
&\quad +2mia\left(
\frac{r^2+a^2}{\Delta}-1
\right)\partial_t
+\ell(\ell+1)
-\partial_r\left(\Delta\partial_r\right)
-\frac{m^2a^2}{\Delta}
\Bigg\}
\sum_{\ell,m}\varphi_{\ell m}Y_{\ell m}.
\end{align}
In obtaining the above expression, we have used the eigenvalue
relations
\begin{equation}
    \hat{L}^2 Y_{\ell m}=\ell(\ell+1) Y_{\ell m}; \ \hat{L}_z Y_{\ell m}=m Y_{\ell m},
\end{equation}
where $\ell$ and $m$ denote the azimuthal and magnetic
quantum numbers, respectively. We next introduce the tortoise
coordinate, which will be useful in the subsequent calculations
\begin{equation}\label{eq:tortoise co}
    \frac{dr_{*}}{dr}=\frac{r^2+a^2}{\Delta}=\frac{1}{f(r)}.
\end{equation}
We now consider the near-horizon limit of the action 
\eqref{eq:action after}. As the event horizon is approached, 
$r\rightarrow r_+$, we have $f(r)\rightarrow 0$. In this limit, 
the dominant contributions to the action \eqref{eq:action after} give
\begin{equation}
\begin{aligned}
S[\varphi]=\;&\int dtdrd\theta d\phi\,\sin\theta
\sum_{\ell',m'}\varphi_{\ell' m'}^{*}Y_{\ell' m'}^{*}
\Bigg[(r^2+a^2)\partial_t^2+2ima\partial_t\\
&\quad-\partial_{r_*}(r^2+a^2)\partial_{r_*}
-\frac{m^2a^2}{r^2+a^2}\Bigg]
\sum_{\ell,m}\varphi_{\ell m}Y_{\ell m}.
\end{aligned}
\end{equation}
Returning to the radial coordinate $r$, we can express the 
resulting effective action as
\begin{equation}
\label{eq:effective action}
\begin{aligned}
S[\varphi]=\sum_{\ell,m}\int dt dr\,(r^2+a^2)\varphi_{\ell m}^{*}
\Bigg[&
\frac{(r^2+a^2)}{\Delta}
\left(\partial_t+\frac{ima}{r^2+a^2}\right)^2-\partial_r\frac{\Delta}{r^2+a^2}\partial_r
\Bigg]\varphi_{\ell m}.
\end{aligned}
\end{equation}
Here, we have used the orthogonality relation of the spherical 
harmonics,
\begin{equation}
\int d\theta\,d\phi\,\sin\theta\,
Y_{\ell m}Y_{\ell' m'}^{*}
=\delta_{\ell,\ell'}\delta_{m,m'}.
\end{equation}
From the effective action in Eq.~\eqref{eq:effective action}, 
we find that $\varphi_{\ell m}$ can be treated as a 
$(1+1)$-dimensional complex scalar field in a two-dimensional 
spherically symmetric spacetime, described by the metric 
$g^{\mu\nu}$, with a dilaton $\Phi$ and a $U(1)$ gauge field:
\begin{equation}
    {\Phi}=r^2+a^2,
\end{equation}
\begin{equation}
     g^{tt}=-f(r),g^{rr}=\frac{1}{f(r)},g^{rt}=0,
\end{equation}
\begin{equation}
    U_{t}=-\frac{a}{r^2+a^2},U_r=0.
\end{equation}
The effective $U(1)$ gauge field arises from the isometry along 
the azimuthal $\phi$ direction, with the corresponding 
two-dimensional field $\varphi_{\ell m}$ carrying a charge $m$. 
Consequently, the associated gauge potential is
\begin{equation}
    A_{t}=m U_t=-\frac{ma}{r^2+a^2}, \  A_r=0.
\end{equation}

Thus, by employing this dimensional reduction procedure, we 
obtain an effective two-dimensional spherically symmetric 
description of the quantum fields in the near-horizon region 
$r\sim r_+$ of the four-dimensional, axisymmetric Kerr spacetime 
\eqref{eq:Kerr_metric}. The resulting effective metric is
\begin{equation}\label{eq: effective metric}
    ds_{\text{eff}}^2=-f(r)dt^2+\frac{dr^2}{f(r)}.
\end{equation}
As a consistency check, the Hawking temperature can be obtained 
from the surface gravity of the effective metric \eqref{eq: effective metric}, and is given by
\begin{equation}
    T_{H}=\frac{f'(r)}{4\pi}=\frac{r_+-r_-}{4\pi(r_+^2+a^2)}.
\end{equation}
The Hawking temperature obtained from the effective metric \eqref{eq: effective metric} agrees with the result in Eq.~\eqref{eq:Hawking temp}. This agreement supports the use of the effective two-dimensional description for calculating 
the entanglement entropy in the following sections.\footnote{An alternative approach to dimensional reduction involves introducing a locally non-rotating coordinate system; see \cite{Murata:2006pt} for details.} In the extremal limit, $r_+=r_-$, the near-horizon geometry reduces to AdS$_2$, in agreement with the result of \cite{Bardeen:1999px}.

\section{Entanglement Entropy: theoretical setup}\label{sec3}

In this section, we present the framework for calculating the entanglement entropy of quantum matter fields in a two-dimensional spacetime, considering different choices of cutoff surfaces. Our analysis is carried out in the ``in'' vacuum state \cite{Fabbri:2005mw,Gan:2022jay,Qu:2024sby}, with the Kerr 
geometry serving as the background spacetime. Since the quantum state is pure on the Cauchy slice $I\cup B\cup R$, the entanglement entropies of complementary regions satisfy
\begin{equation}
    S_{\text{ent}}(B)=S_{\text{ent}}(I\cup R).
\end{equation}
Consequently, as follows from Eqs.~\eqref{eq:sgen for BH} and \eqref{eq:sgen for R}, we have $S_{BH}=S(R)$. For convenience, we will therefore focus primarily on calculating $S_{\text{ent}}(B)$ in the following analysis.

\subsection{Cutoff Surface away from the horizon}\label{subsec3}

When the cutoff surface $A$ is sufficiently far from the black hole horizon, we consider the massless scalar fields in the $s$-wave approximation.\footnote{The $s$-wave approximation retains the lowest angular momentum mode of the matter field, thereby reducing the problem to an effective two-dimensional description.} In this regime, the matter contribution to the entanglement entropy can be evaluated using the vacuum result for a two-dimensional conformal field theory (CFT) in flat spacetime. For an interval with endpoints $A$ and $I$, it is given by \cite{Calabrese:2004eu,Calabrese:2009qy,Casini:2009sr}
\begin{equation}
    S_{\text{ent}}=\frac{c}{3}\ln\left(d(A,I)\right),
\end{equation}
where $c$ is the central charge and $d(A,I)$ denotes the distance between the two endpoints. In light-cone coordinates $(x^+,x^-)$, with the flat-space line element $ds^2=-dx^+dx^-$, this distance is expressed as
\begin{equation}
d(A,I) = \sqrt{ \left[ x^+(A) - x^+(I) \right] \,
\left[ x^-(I) - x^-(A) \right] }.
\end{equation}
The vacuum state is defined with respect to the coordinates $(x^+,x^-)$, such that the vacuum expectation values of the normal-ordered stress-tensor components vanish:
\begin{equation}
\langle 0| : T_{x^{+}x^{+}}:|0\rangle
=
\langle 0| : T_{x^{-}x^{-}}:|0\rangle=0.
\end{equation}
For a general two-dimensional curved spacetime, the metric can be expressed in conformal coordinates as
\begin{equation}
    ds^2 = -e^{2\rho(x^+,x^-)} dx^+ dx^-,
\end{equation}
where $\rho(x^+,x^-)$ is the conformal factor. For an arbitrary quantum state $|\Psi\rangle$, the expectation values of the normal-ordered stress tensor are characterized by two state-dependent functions $t_\pm(x^\pm)$
\begin{equation}
    \langle \Psi | : T_{\pm\pm}(x^{\pm}):|\Psi\rangle
    =-\frac{\hbar}{12\pi}t_{\pm}(x^{\pm}).
\end{equation}
Under a Weyl transformation $g\rightarrow\Omega^2g$, the entanglement entropy transforms as
\cite{Almheiri:2019psf}
\begin{equation}
    S_{\Omega^2 g}=S_g-\frac{c}{6}
    \sum_{\text{endpoints}}\ln(\Omega),
\end{equation}
where the sum runs over the endpoints of the interval under consideration. Using $\Omega^{-2}=e^{2\rho}$ and incorporating the flat-space result, we obtain the matter contribution to the generalized entropy in a two-dimensional curved spacetime \cite{Gautason:2020tmk}\footnote{Throughout this paper, we use the subscript $I$ to denote coordinates associated with $\partial I$.}
\begin{equation}\label{eq:sent formula cut off far}
    S_{\text{ent}}
    =\frac{c}{6}\ln\left(d(A,I)^2e^{\rho_A}e^{\rho_I}\right)_{t_{\pm}=0}
    =\frac{c}{12}\ln\left(d(A,I)^4e^{2\rho_A}e^{2\rho_I}\right)_{t_{\pm}=0}.
\end{equation}
Here, $\rho_A$ and $\rho_I$ denote the conformal factors evaluated at the cutoff surface and the island boundary, respectively. The condition $t_\pm=0$ specifies the vacuum with respect to the chosen null coordinates.\footnote{In the present work, we consider the ``in'' vacuum associated with the one-sided Kerr black hole formed through gravitational collapse. Accordingly, we choose the null coordinates $x^+=u_{\mathrm{in}}$ and $x^-=v$, which define the in-vacuum state $|\text{in}\rangle$. Thus, the condition $t_\pm=0$ is imposed with respect to these coordinates, and the conformal factor is evaluated as
$(e^{2\rho})_{t_\pm=0}
=e^{2\rho(u_{\mathrm{in}},v)}$.}
The UV cutoff dependence has been omitted, as it can be
absorbed into the renormalization of the gravitational
coupling $G$
\cite{Gautason:2020tmk,Almheiri:2019psf,Nelson:1994na,
Alishahiha:2020qza,Kim:2021gzd,Hashimoto:2020cas}.
For a more detailed discussion of these topics, see
\cite{Gan:2022jay,Qu:2024sby}.

\subsection{Cutoff surface close to the horizon}\label{subsec4}

When the characteristic size of a region is much smaller than the local curvature scale, the entanglement entropy of matter fields can be approximated by its flat-spacetime vacuum expression. This local approximation is applicable even in curved backgrounds, including the near-horizon region of a black 
hole. Accordingly, the renormalized matter contribution to the 
entanglement entropy takes the form 
\cite{Casini:2005zv,Hashimoto:2020cas,Qu:2024sby,Gan:2022jay}
\begin{equation}\label{eq:entropy near formula}
S_{\text{ent}} = -K c \frac{A}{L^2},
\end{equation}
where $K$ is a constant associated with the renormalization 
procedure, $c$ denotes the central charge of the field theory, 
$A$ is the area of the cutoff surface, and $L$ represents the 
geodesic distance between the cutoff surface $A$ and the island 
boundary $\partial I$.When both $\partial I$ and the cutoff surface $A$ lie close to the horizon, the geodesic distance can be approximated as 
\cite{Hashimoto:2020cas,Qu:2024sby,Gan:2022jay}
\begin{equation}
L = \int \sqrt{-e^{2\rho(x^+,x^-)} dx^+ dx^-} 
\end{equation}
\begin{equation}
    \approx \sqrt{\left[x^+(A) - x^+(I)\right]\left[x^-(I) - x^-(A)\right]e^{\rho_A} e^{\rho_I} } 
\end{equation}
\begin{equation}
    \approx\sqrt{ d(A, I)^2 e^{\rho_A} e^{\rho_I}.}
\end{equation}
Here, $d(A,I)$ denotes the coordinate separation between the cutoff surface and the island boundary $\partial I$. The above approximation is valid when their separation is sufficiently small compared with the local curvature scale. It therefore provides a useful local description for evaluating the matter entanglement 
entropy near the black hole horizon.
\section{Non-extremal Kerr black hole}\label{sec 4}
\subsection{Connecting condition}\label{subsec5}
We consider a one-sided non-extremal Kerr black hole formed by the collapse of a spherical null shell at $v=v_0$. The spacetime consists of two regions, corresponding to the geometries before and after the shell collapse. The matching of these two regions determines the connecting condition, which plays an important role in describing the near-horizon geometry and its subsequent
analysis. The corresponding spacetime configuration is illustrated in the Penrose diagram in Fig.~\ref{fig:Penrose_diagram_non-extremal}.\\
Before the collapse, in the ``in'' region ($v<v_0$), the spacetime is described by the Minkowski metric
\begin{equation}\label{Minkowski reg}
    ds^2 = -du_{\text{in}} dv + r_{\text{in}}^2 d\Omega^2,
\end{equation}
where $u_{\text{in}}$ and $v$ are the outgoing and ingoing null coordinates, respectively, and $r_{\text{in}}$ denotes the radial coordinate in this region.\\
After the collapse, in the ``out'' region ($v>v_0$), the geometry is described by the effective two-dimensional Kerr metric.\footnote{Here, we work with the effective two-dimensional description of the Kerr geometry, in which the angular dependence has been integrated out and the effects of rotation
are encoded in the parameter $a$.} In double-null coordinates $(u,v)$, the metric takes the form
\begin{equation}
    ds^2=-\frac{(r-r_{+})(r-r_{-})}{r^2+a^2}dudv,
\end{equation}
where the null coordinates are related by
\begin{equation}\label{eq:Kruskal co trans}
    u_{in}=t_{in}-r_{in} , u=t-r_{*} , v=t_{in}+r_{in}=t+r_{*},
\end{equation}
and $r_*$ is the tortoise coordinate associated with the effective two-dimensional metric introduced earlier \eqref{eq: effective metric}. It is given by
\begin{equation}\label{eq:non-extre tor}
    r_{*}(r)=\int\frac{dr}{f(r)}=r+\frac{a^2+r_+^2}{r_{+}-r_{-}}\ln\Bigg|\frac{{r-r_{+}}}{r_{+}+r_{-}}\Bigg|-\frac{a^2+r_-^2}{r_{+}-r_{-}}\ln\Bigg|\frac{{r-r_{-}}}{r_{+}+r_{-}}\Bigg|.
\end{equation}
Matching the two spacetime regions across the collapsing shell at $v=v_0$ gives the connecting condition between the outgoing null coordinates $u_{\text{in}}$ and $u$:
\begin{equation}\label{connecting condition}
    u=u_{in}+\frac{2(r_{-}^2+a^2)}{r_+-r_-}\ln\Bigg|\frac{v_0-u_{in}-2r_-}{2(r_{+}+r_{-})}\Bigg|-\frac{2(r_{+}^2+a^2)}{r_+-r_-}\ln\Bigg|\frac{v_0-u_{in}-2r_+}{2(r_{+}+r_{-})}\Bigg|.
\end{equation}
This relation is particularly useful in the late-time limit. As $u\rightarrow+\infty$, the incoming null coordinate
approaches
\begin{equation}
    u_{\text{in}}\approx-2(r_{+}+r_{-})e^{-\frac{u(r_+-r_-)}{2(a^2+r_+^2)}}-2r_+ +v_0,
\end{equation}
which can equivalently be written as
\begin{equation}
    u_{\text{in}}\approx-2(r_{+}+r_{-})e^{-\kappa u}-2r_+ +v_0.
\end{equation}
To describe the geometry across the event horizon, we introduce
the Kruskal coordinates. They are defined separately in the exterior and interior regions as
\begin{equation}\label{eq:tor outside}
    U=-e^{-\kappa u} , V=e^{\kappa v}
\end{equation}
outside the horizon, and
\begin{equation}\label{eq:tor inside}
    U=e^{-\kappa u} , V=e^{\kappa v}
\end{equation}
inside the horizon, where $\kappa$ is the surface gravity defined in Eq.~\eqref{Surface gravity}. In these coordinates, the line element becomes
\begin{equation}
    ds^2=\frac{f(r)dU dV}{\kappa^2 U V},
\end{equation}
with $ f(r)=\frac{(r-r_{+})(r-r_{-})}{r^2+a^2}.$
Using the late-time relation between $u_{\text{in}}$ and $u$, we can also express the incoming null coordinate in terms of the Kruskal coordinate $U$ as
\begin{equation}
    u_{in}\approx2(r_{+}+r_{-})U-2r_++v_0,
\end{equation}
which gives
\begin{equation}
    \frac{dU}{du_{in}}\approx \frac{1}{2(r_{+}+r_{-})} , \frac{dV}{dv}=\kappa V.
\end{equation}
We can now express the metric throughout the spacetime in conformal double-null form, $ds^2=-e^{2\rho(u_{in},v)}du_{in}dv$. In the ``in'' region,
where $v<v_0$, the conformal factor is $e^{2\rho(u_{in},v)}=1$, corresponding to the flat spacetime geometry. In the ``out'' region, the conformal factor is obtained from the Kruskal-coordinate expression $e^{2\rho(U,V)}$ through the coordinate transformation:
\begin{equation}
    e^{2\rho(u_{in},v)}=e^{2\rho(U,V)}\frac{dU}{du_{in}}\frac{dV}{dv}=-\frac{f(r)}{2(r_{+}+r_{-})\kappa U}.
\end{equation}
This provides the conformal factor in the coordinates $(u_{\text{in}},v)$, which will be used in the subsequent calculation of the entanglement entropy.

\subsection{Cutoff surface away from the horizon}\label{subsec6}

The location of the cutoff surface plays an important role in determining the matter contribution to the entanglement entropy. We first investigate the non-extremal rotating Kerr black hole when the cutoff surface is situated away from the event horizon. We consider both the island and no-island configurations, and begin with the former.

\subsubsection{With island}\label{subsubsec1}

\begin{figure}[h]
    \centering 
    \includegraphics[width=0.5\textwidth]{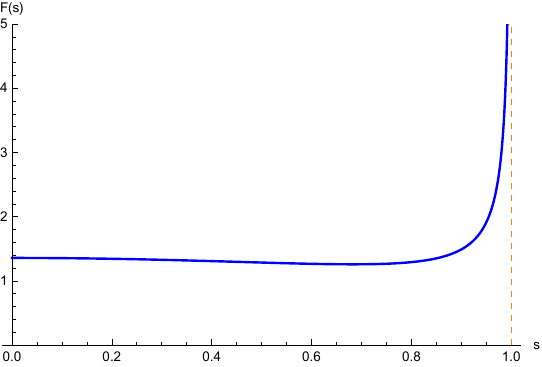}
    \caption{The variation of $F(s)$ with $s=\frac{a}{M}$. The function remains nearly constant over most of the parameter range and increases sharply as $s$ approaches unity.}
    \label{fig:Ks_vs_s}
\end{figure}

In the presence of an island, the generalized entropy is given by
\begin{equation}\label{eq:s gen new}
    S_{\text{gen}}=S_{\text{gravity}}+S_{\text{ent}},
\end{equation}
where 
\begin{equation}\label{eq:S gravity}
    S_{\text{gravity}}=\frac{\pi (r_{I}^2+a^2)}{G_{N}}
\end{equation}
is the gravitational contribution associated with the area of the island boundary $\partial I$.\\
Since the island is considered sufficiently far from the horizon, we employ the $s$-wave approximation for the matter fields. We work in the ``in'' vacuum state $|\text{in}\rangle$, defined in the pre-collapse Minkowski region ($v<v_0$) with respect to the coordinates $(u_{\text{in}},v)$. In this coordinate system, the vacuum expectation values of the normal-ordered stress tensor vanish \cite{Qu:2024sby,Fabbri:2005mw}. Consequently, the state-dependent functions satisfy $t_{\pm}=0$, and the corresponding conformal factor is therefore
\begin{equation}
    (e^{2\rho})_{t_\pm=0}=e^{2\rho(u_{in},v)}.
\end{equation}
This choice of quantum state is an important feature of our analysis and distinguishes the present setup from previous studies employing different vacuum prescriptions.\\
The position of the island relative to the event horizon determines the appropriate relation between its radial coordinate and the Kruskal coordinates. We first consider the case in which the island lies outside the horizon. In this case, we have
\begin{equation}
    U_IV_I=-e^\frac{{(r_+-r_-) r_{I}}}{r_+^2+a^2}\left(\frac{r_{I}-r_+}{r_++r_-}\right)\left(\frac{r_{I}-r_-}{r_++r_-}\right)^{-\frac{a^2+r_-^2}{a^2+r_+^2}}.
\end{equation}
Taking the island to be close to the event horizon, $r_I\approx r_+$, this expression becomes
\begin{equation}
    U_IV_I \approx-e^\frac{{(r_+-r_-) r_{+}}}{r_+^2+a^2}\left(\frac{r_{I}-r_+}{r_++r_-}\right)\left(\frac{r_{+}-r_-}{r_++r_-}\right)^{-\frac{a^2+r_-^2}{a^2+r_+^2}}.
\end{equation}
Solving for the island position, we obtain
\begin{equation}
    r_{I}\approx r_{+}-(r_++r_-)e^{-\frac{(r_+-r_-)r_+}{r_+^2+a^2}}\left(\frac{r_+-r_-}{r_++r_-}\right)^{\frac{a^2+r_-^2}{a^2+r_+^2}}U_{I}V_{I}=r_{+}-\zeta U_{I}V_{I},
\end{equation}
where
$\zeta=(r_++r_-)e^{-\frac{(r_+-r_-)r_+}{r_+^2+a^2}}\left(\frac{r_+-r_-}{r_++r_-}\right)^{\frac{a^2+r_-^2}{a^2+r_+^2}}$.
We now turn to the configuration in which the island is located inside the event horizon. The corresponding relation between the Kruskal coordinates and the island position is
\begin{equation}
    U_IV_I=e^\frac{{(r_+-r_-) r_{I}}}{r_+^2+a^2}\left(\frac{r_{+}-r_I}{r_++r_-}\right)\left(\frac{r_{I}-r_-}{r_++r_-}\right)^{-\frac{a^2+r_-^2}{a^2+r_+^2}}.
\end{equation}
Following the same near-horizon approximation, we obtain
\begin{equation}\label{eq:island position far from horizon}
    r_I \approx r_+-\zeta U_IV_I .
\end{equation}
Thus, the near-horizon relation between the island position and the Kruskal coordinates takes the same form for both configurations.\\
Using the $s$-wave approximation together with the entanglement entropy formula derived earlier, the matter contribution can be written as
\begin{equation}
    S_{\text{ent}}=\frac{c}{12}\ln \Bigg[[u_{\text{in}}(A)-u_{\text{in}}(I)]^2[v_I-v_A]^2\frac{f(r_A)f(r_I)}{\kappa^2 U_IU_A}\Bigg].
\end{equation}
Substituting the near-horizon relation for $r_I$ and the corresponding coordinate transformations, we obtain
\begin{equation}\label{Entg formula}
     S_{\text{ent}}\approx \frac{c}{12}\ln \Bigg[(U_A-U_I)^2 \left(\ln \frac{V_I}{V_A}\right)^2 \frac{V_I\zeta\left(r_+-r_--U_IV_I\zeta\right)}{\kappa^4\left((r_+-U_IV_I\zeta)^2+a^2\right) U_A}\Bigg],
\end{equation}
where we have used the approximation $f(r_A)\approx1$ for $r_A\gg r_+$. Since the island boundary is assumed to lie close to the event horizon, we take $U_I\rightarrow0$. Expanding Eq.~\eqref{Entg formula} to first order in $U_I$,
we find
\begin{equation}
\begin{aligned}
S_{\text{ent}}=\frac{c}{12}\Bigg\{&
\ln\Bigg|\frac{U_A V_I \zeta (r_+-r_-)}
{\kappa^4 (r_+^2+a^2)}
\left(\ln \frac{V_I}{V_A}\right)^2\Bigg|\\
&+\frac{r_+^2-2r_+r_--a^2}
{(r_+-r_-)(r_+^2+a^2)}
U_I V_I \zeta
-2\frac{U_I}{U_A}
+\mathcal{O}(U_I^2)
\Bigg\}.
\end{aligned}
\end{equation}
We next evaluate the gravitational contribution by substituting the near-horizon expression for $r_I$:
\begin{equation}
\begin{aligned}
S_{\text{gravity}}
=\;&\frac{\pi}{G}(r^2_I+a^2)
=\frac{\pi}{G}\left[\left(r_+-U_IV_I\zeta\right)^2+a^2\right]\\
&=\frac{\pi( r_+^2+a^2)}{G}
-\frac{2\pi U_I V_I\zeta r_+}{G}
+\mathcal{O}(U_I^2).
\end{aligned}
\end{equation}
Combining the matter and gravitational contributions, the generalized entropy becomes
\begin{equation}\label{S gen}
\begin{split}
    S_{\text{gen}} &= \frac{\pi \,( r_+^2+a^2)}{G} 
    + \frac{c}{12} \ln \Bigg| 
        \frac{U_A V_I\zeta (r_+ - r_-)}{\kappa^4 \, (r_+^2+a^2)}
        \left( \ln \frac{V_I}{V_A} \right)^2 
    \Bigg| \\
    &\quad + \frac{c}{12} U_I 
    \Bigg[
        V_I \zeta 
        \left( \frac{r_+^2 - 2r_+r_--a^2}{(r_+-r_-)(r_+^2+a^2)} - \frac{24 \pi r_+}{cG} \right) 
        - \frac{2}{U_A}
    \Bigg]  + \mathcal{O}(U_I^2).
\end{split}
\end{equation}
The location of the quantum extremal surface is determined by extremizing the generalized entropy with respect to the island coordinates. First, varying Eq.~\eqref{S gen} with respect to $U_I$, we obtain
\begin{equation}\label{eq:series}
\frac{\partial S_{\text{gen}}}{\partial U_I}
\approx
V_I \zeta
\left(
\frac{r_+^2 - 2 r_+ r_- - a^2}
{(r_+ - r_-)(r_+^2 + a^2)}
-
\frac{24 \pi r_+}{c G}
\right)
-
\frac{2}{U_A}
= 0.
\end{equation}
Solving this extremality condition for $V_I$ gives
\begin{equation}\label{eq:VI}
    V_I = \frac{2 c G (r_+-r_-)(r_+^2+a^2)}{U_{A}\zeta \left[ c G (r_+^2 -  2r_+r_--a^2) - 24 \pi r_+(r_+-r_-)(r_+^2+a^2) \right] }.
\end{equation}
We then impose the second extremality condition by varying the generalized entropy with respect to $V_I$:
\begin{equation}\label{eq:derivative}
    \frac{\partial S_{\text{gen}}}{\partial V_I} \approx \frac{  1 +\frac{2}{ \ln \frac{V_I}{V_A}} }{ V_I }+{ U_I\zeta\left(\frac{r_+^2-2r_+r_--a^2}{(r_+-r_-)(r_+^2+a^2)}-\frac{24\pi r_+}{cG}\right)} =0.
\end{equation}
 The term $\frac{2}{\ln(V_I/V_A)}$ is subleading and can be neglected due to its smallness. Using Eq.~\eqref{eq:VI}, the second extremality condition
then yields
\begin{equation}\label{eq:UI}
    U_I=-\frac{U_A}{2}.
\end{equation}
Since $U_A<0$, thus $U_I>0$, which means the island boundary $\partial I$ lies inside the event horizon. This result differs from the location obtained in the Hartle-Hawking vacuum, where the island is found outside the event horizon \cite{Wang:2024itz,Yu:2021rfg}. The choice of quantum state therefore has an important effect on the resulting island configuration.\\
Furthermore, combining the extremality conditions gives
\begin{equation}\label{UIVI}
    U_IV_I=-\frac{ c G (r_+-r_-)(r_+^2+a^2)}{\zeta \left[ c G (r_+^2 -  2r_+r_--a^2) - 24 \pi r_+(r_+-r_-)(r_+^2+a^2) \right] }\approx\frac{cG}{24 \pi\zeta \ r_+}<<1,
\end{equation}
where we have employed the semiclassical approximation, in which quantum contributions are subdominant to the geometric scales of the horizon:
$\frac{cG}{\zeta r_+}<<1$. Consequently, as $U_IV_I\rightarrow0$, we have $r^{*}_{I}\rightarrow-\infty$, confirming the consistency of the near-horizon approximation $r_I\approx r_+.$\\
To examine the dependence of the island position on the rotation parameter, we substitute Eq.~\eqref{roots} into Eq.~\eqref{UIVI} and set $a=sM$, where $s$ is a dimensionless parameter.\footnote{The parameter $a=\frac{J}{M}$ has dimensions of length in natural units, as can be seen from Eq.~\eqref{eq:parameters}. Thus, we write $a=sM$, where $s$ is dimensionless.} The result takes the form
\begin{equation}
    U_IV_I=\frac{c G}{48\pi M^2}F(s),
\end{equation}
where
\begin{equation}
    F(s)=\frac{(1-s^2)^{-\frac{\left(1-\sqrt{1-s^2}\right)^2+s^2}{2\left(\left(1+\sqrt{1-s^2}\right)^2+s^2\right)}}}{1+\sqrt{1-s^2}}\text{exp}\left[\frac{2\sqrt{1-s^2}\left(1+\sqrt{1-s^2}\right)}{\left(1+\sqrt{1-s^2}\right)^2+s^2}\right].
\end{equation}
For $a=0$, we obtain $F(s)=e/2$, corresponding to the Schwarzschild limit. As $s$ approaches unity, which corresponds to the near-extremal Kerr regime, $F(s)$ grows without bound. This indicates that the near-horizon approximation used to determine the island position may cease to be reliable sufficiently close to extremality, since $U_IV_I$ is no longer guaranteed to remain small.
The dependence of $F(s)$ on the dimensionless rotation parameter is illustrated in Fig.~\ref{fig:Ks_vs_s}.
Finally, substituting the extremal values of $U_I$ and $V_I$ back into Eq.~\eqref{S gen}, we obtain
\begin{align}\label{eq:sgen final}
S_{\text{gen}} 
=&\; \frac{\pi (r_+^2 + a^2)}{G} 
+ \frac{c}{12} \ln \Bigg| 
\frac{2 c G (r_+ - r_-)^2}
{\kappa^4 \left[ c G (r_+^2 - 2r_+ r_- - a^2) 
- 24 \pi r_+ (r_+ - r_-) (r_+^2 + a^2) \right]} \notag \\
&\qquad 
\left(
\ln \frac{2 c G (r_+ - r_-) (r_+^2 + a^2)}
{\zeta U_A V_A 
\left[ c G (r_+^2 - 2r_+ r_- - a^2) 
- 24 \pi r_+ (r_+ - r_-) (r_+^2 + a^2) \right]}
\right)^2
\Bigg|
+ \mathcal{O}(G)
\notag \\
\approx&\; \frac{\pi (r_+^2 + a^2)}{G}
= S_{BH}.
\end{align}
In the final approximation, we have again used $\frac{cG}{\zeta r_+}<<1$. Thus, the generalized entropy approaches the Bekenstein-Hawking entropy at late times. The island contribution therefore prevents the radiation entropy from growing indefinitely and reproduces the expected saturation at the black-hole entropy, which decreases as the black hole evaporates.

\subsubsection{Without island}\label{subsubsec2}

We now consider the configuration without an island, for which we set $r_I=0$.\footnote{For $r_I=0$, the gravitational contribution given by Eq.~\eqref{eq:S gravity} is not exactly zero but remains finite due to the rotation parameter $a$. However, this contribution is independent of the radiation time $t_A$ and therefore enters the generalized entropy only as an additive constant. Since it does not affect the time evolution of the radiation entropy, it will be omitted in the following analysis.} Since this point lies in the pre-collapse Minkowski region ~\eqref{Minkowski reg}, the corresponding coordinates and conformal factor satisfy $u_{in}(I)=t_I$ and $e^{2\rho(u_{in},v)(I)}=1$, respectively. Consequently, the generalized entropy takes the form
\begin{equation}
    S_{\text{gen}}=\frac{c}{12}\ln \Bigg|\left(2U_A+v_0-2r_+-t_I\right)(v_I-v_A)^2 \frac{f(r_A)}{2\kappa U_A}\Bigg|.
\end{equation}
Using the corresponding coordinate transformations, we can express the generalized entropy  as
\begin{equation}
\begin{aligned}
    S_{\text{gen}}
    =\;&\frac{c}{6}\ln\Bigg|-2e^{-\kappa(t_A-r_A^*)}+v_0-2r_+-t_I\Bigg|\\
    &+\frac{c}{6}\ln\Bigg|t_{I}-t_A-r_A^*\Bigg|
    +\frac{c}{12}\ln\Bigg|\frac{f(r_A)}{2\kappa}\Bigg|
    -\frac{c}{12}\ln\Bigg|-e^{-\kappa(t_A-r_A^*)}\Bigg|.
\end{aligned}
\end{equation}
We are particularly interested in the late-time behavior of the radiation entropy. In the regime $t_A\gg r_A\gg r_+$, the leading contribution to the
generalized entropy becomes
\begin{equation}\label{eq:withot island result}
    S_{\text{gen}}\approx\frac{c}{12}\kappa t_A,
\end{equation}
where subleading logarithmic contributions, including the term proportional to $\ln(t_A)$, have been neglected. Thus, in the absence of an island, the entropy of Hawking radiation increases linearly with time at late times.
This unbounded growth is consistent with the conventional semiclassical prediction of Hawking radiation.

\subsection{Cutoff surface close to the horizon}\label{subsec6}
In this section, we study a one-sided rotating Kerr black-hole formed by the collapse of a null shell, considering the corresponding ``in'' vacuum state. We focus on the case in which the cutoff surface is located close to the black hole horizon and follow an analytical approach similar to that of Ref.~\cite{Qu:2024sby}, which studied one-sided charged Reissner--Nordström black holes. Since the region of interest is much smaller than the characteristic curvature scale, the matter-field entropy can be approximated by its flat-space expression. This approximation does not depend on the
details of the spacetime geometry or the choice of quantum state.

\subsubsection{With island}

We now turn to the configuration in which an island is present. The generalized entropy receives contributions from the gravitational entropy and the matter sector. The gravitational contribution is given by
\begin{equation}
    S_{\text{gravity}}
    =\frac{\pi(r_I^2+a^2)}{G}.
\end{equation}
Since both the island boundary $\partial I$ and the cutoff surface $A$ are taken to lie close to and outside the horizon, the effective distance between them can be approximated as
\begin{equation}
    L\approx
    \sqrt{d(A,I)^2 e^{\rho_A}e^{\rho_I}}.
\end{equation}
Using the near-horizon relations, this becomes
\begin{equation}
    =\frac{1}{\kappa}\sqrt{(U_A-U_I)(V_I-V_A)\sqrt{f(r_A)f(r_I)}e^{-\kappa(r_A^*+r_I^*)}}.
\end{equation}
Substituting this expression into the matter entropy formula in Eq.~\eqref{eq:entropy near formula}, we obtain
\begin{equation}
    S_{\text{gen}}
    =\frac{\pi(r_I^2+a^2)}{G}
    -Kc\frac{A}{L^2}.
\end{equation}
Consequently, the generalized entropy  can be written as
\begin{equation}
\begin{aligned}
    S_{\text{gen}}
    ={}&\frac{\pi(r_I^2+a^2)}{G}
    -\frac{
    4\pi Kc\,r_A^2\kappa^2
    e^{\kappa(r_A^*+r_I^*)}
    }{
    (U_A-U_I)(V_I-V_A)
    \sqrt{f(r_A)f(r_I)}
    }.
\end{aligned}
\end{equation}
Using the explicit form of the blackening factor \eqref{eq:tortoise co}, we arrive at
\begin{equation}\label{s gen case 2}
\begin{aligned}
S_{\text{gen}}
={}&\frac{\pi(r_I^2+a^2)}{G}
\\[4pt]
&-\frac{
e^{\kappa(r_A^*+r_I^*)}
}{
\left(-e^{-\kappa(t_A-r_A^*)}
+e^{-\kappa(t_I-r_I^*)}\right)
\left(e^{\kappa(t_I+r_I^*)}
-e^{\kappa(t_A+r_A^*)}\right)
}
\\[-1pt]
&\qquad\times
\frac{4\pi cK r_A^2\kappa^2\sqrt{a^2+r_I^2}\sqrt{a^2+r_A^2}}{
\sqrt{(r_A-r_-)(r_I-r_-)
(r_A-r_+)(r_I-r_+)}
}.
\end{aligned}
\end{equation}
We first extremize the generalized entropy with respect to the island time coordinate $t_I$. This gives
\begin{equation}
\begin{aligned}
\frac{\partial S_{\text{gen}}}{\partial t_I}
={}&-\frac{
 e^{(t_A+t_I)\kappa}
\left(e^{2t_A\kappa}-e^{2t_I\kappa}\right)
}{
\left(e^{(-r_A^*+t_A)\kappa}
-e^{(-r_I^*+t_I)\kappa}\right)^2
\left(e^{(r_A^*+t_A)\kappa}
-e^{(r_I^*+t_I)\kappa}\right)^2}
\\
&\quad\times
\frac{4\pi cK r_A^2\kappa^3\sqrt{a^2+r_I^2}\sqrt{a^2+r_A^2}}{
\sqrt{(r_A-r_-)(r_I-r_-)
(r_A-r_+)(r_I-r_+)}
}
=0.
\end{aligned}
\end{equation}
The corresponding solutions are
\begin{equation}
    t_I=t_A+\frac{i\pi C_1}{\kappa},
    \qquad C_1\in\mathbb Z.
\end{equation}
Restricting to real-valued solutions, we take
\begin{equation}
    t_I=t_A.
\end{equation}
Substituting this relation into Eq.~\eqref{s gen case 2}, the generalized entropy reduces to
\begin{equation}\label{S Gen for t_I=t_A}
S_{\text{gen}}=\frac{\pi (a^2 + r_I^2)}{G} - 
\frac{4\pi c  K  r_A^2 e^{\kappa (r_A^* + r_I^*)}   \kappa^2\sqrt{(a^2+r_{A}^2)}\sqrt{(a^2+r_I^2)}}{
\left(e^{\kappa r_A^*} - e^{\kappa r_I^*}\right)^2 
\sqrt{(r_A - r_-) (r_I - r_-) (r_A - r_+) (r_I - r_+)}
}.
\end{equation}
We now use the near-horizon parametrization
\begin{equation}\label{eq:island loc cut near}
    r_A=r_+(1+\alpha),
    \qquad
    r_I=r_+(1+\beta),
\end{equation}
where $0<\beta<\alpha\ll1$. In this regime, both the cutoff surface and the island boundary are located close to the outer horizon. For a cutoff surface close to the horizon, no physical solution exists for $\partial I$ to lie inside the horizon, as explained in appendix \ref{app1}.\\
Expanding Eq.~\eqref{S Gen for t_I=t_A} in powers of $\beta$ and retaining terms up to first order, we obtain
\begin{equation}\label{eq:sgen with beta}
\begin{split}
S_{\text{gen}}={}&
\frac{\pi(a^2+r_+^2)}{G}
\\
&-\frac{
4\kappa^2\pi cK r_+e^{-\kappa\alpha r_+}
(1+\alpha)^2
\sqrt{a^2+r_+^2(1+\alpha)^2}
\sqrt{a^2+r_+^2}
}{
\alpha(r_+-r_-)
\sqrt{1+\dfrac{r_+\alpha}
{2\kappa(r_+^2+a^2)}}
}
\\[3mm]
&-\frac{
2\sqrt{\beta}\,
\left[
4\kappa^2\pi cK r_+
e^{-2\kappa\alpha r_+}
(1+\alpha)^2
\sqrt{a^2+r_+^2(1+\alpha)^2}
\sqrt{a^2+r_+^2}
\right]
}{
\alpha^{3/2}(r_+-r_-)
\sqrt{1+\dfrac{r_+\alpha}
{2\kappa(r_+^2+a^2)}}
}
\\[3mm]
&+\beta\bigg\{
\frac{2\pi r_+^2}{G}
\\
&\quad
-\frac{
4\kappa^2\pi cK r_+
e^{-3\kappa\alpha r_+}
(1+\alpha)^2
\sqrt{a^2+r_+^2(1+\alpha)^2}
\sqrt{a^2+r_+^2}
}{
\alpha^2(r_+-r_-)
\sqrt{1+\dfrac{r_+\alpha}
{2\kappa(r_+^2+a^2)}}
}
\\
&\qquad\times
\left[
\alpha e^{2\alpha\kappa r_+}
\left(
\frac{r_+^2}{r_+^2+a^2}
+\kappa r_+
-\frac{r_+}{4\kappa(r_+^2+a^2)}
\right)
+3
\right]
\bigg\}+\mathcal{O}(\beta^{3/2}).
\end{split}
\end{equation}
The island location is determined by extremizing the generalized entropy with respect to $\beta$. Differentiating Eq.~\eqref{eq:sgen with beta}, we find
\begin{equation}
\begin{aligned}
\frac{\partial S_{\text{gen}}}{\partial\beta}
={}&\frac{2\pi r_+^2}{G}
\\
&-\frac{
4\kappa^2\pi cK r_+
e^{-3\kappa\alpha r_+}
(1+\alpha)^2
\sqrt{a^2+r_+^2(1+\alpha)^2}
\sqrt{a^2+r_+^2}
}{
\alpha^2(r_+-r_-)
\sqrt{1+\dfrac{r_+\alpha}
{2\kappa(r_+^2+a^2)}}
}
\\
&\quad\times
\left[
\alpha e^{2\alpha\kappa r_+}
\left(
\frac{r_+^2}{r_+^2+a^2}
+\kappa r_+
-\frac{r_+}{4\kappa(r_+^2+a^2)}
\right)
+3
\right]
\\
&-\frac{
4\kappa^2\pi cK r_+
e^{-2\kappa\alpha r_+}
(1+\alpha)^2
\sqrt{a^2+r_+^2(1+\alpha)^2}
\sqrt{a^2+r_+^2}
}{
\sqrt{\beta}\,\alpha^{3/2}(r_+-r_-)
\sqrt{1+\dfrac{r_+\alpha}
{2\kappa(r_+^2+a^2)}}
}
=0.
\end{aligned}
\end{equation}
Solving this equation, we obtain
\begin{equation}\label{eq:sol of beta}
\begin{aligned}
\beta
\approx{}&
\frac{
16\alpha\kappa^4c^2G^2K^2r_+^2
e^{-4\kappa\alpha r_+}
(1+\alpha)^4
\left(a^2+r_+^2(1+\alpha)^2\right)
(a^2+r_+^2)
}{
\left[
2\alpha^2r_+^2(r_+-r_-)
\sqrt{1+\dfrac{r_+\alpha}
{2\kappa(r_+^2+a^2)}}
\right]^2
}
\\[3pt]
\approx{}&
\frac{c^2G^2K^2}{4\alpha^3}
\frac{(r_+-r_-)^2}
{r_+^2(a^2+r_+^2)^2}
\ll1.
\end{aligned}
\end{equation}
This result shows that the island boundary $\partial I$ lies close to and just outside the horizon, provided $c^2G^2\ll \dfrac{r_+^2(a^2+r_+^2)^2}{(r_+-r_-)^2}$. Substituting Eq.~\eqref{eq:sol of beta} into Eq.~\eqref{eq:sgen with beta} and expanding the resulting expression to zeroth order in $G$, we obtain
\begin{equation}
\begin{aligned}
S_{\text{gen}}
={}&\frac{\pi(r_+^2+a^2)}{G}
\\
&-\frac{
4\kappa^2\pi cK r_+
e^{-\kappa\alpha r_+}
(1+\alpha)^2
\sqrt{a^2+r_+^2(1+\alpha)^2}
\sqrt{a^2+r_+^2}
}{
\alpha(r_+-r_-)
\sqrt{1+\dfrac{r_+\alpha}
{2\kappa(r_+^2+a^2)}}
}+\mathcal{O}(G)
\\
\approx{}&S_{\text{BH}},
\end{aligned}
\end{equation}
where in obtaining the above result, we have taken into account the condition
$cG\ll \dfrac{r_+(r_+^2+a^2)}{r_+-r_-}$. Thus, for a cutoff surface located close to the horizon, the island configuration ensures that the entropy of Hawking radiation remains bounded by the Bekenstein-Hawking entropy at late times.

\subsubsection{Without island}

For the islandless configuration, the matter entropy $S_{\text{ent}}$ must be evaluated using Eq.~\eqref{eq:sent formula cut off far}. This is because
the island endpoint is set to $r_I=0$, which lies far from the near-horizon cutoff surface (See the right panel of Fig.~\ref{fig:Penrose_diagram_non-extremal} for the Penrose diagram of the no-island configuration). For a large macroscopic black hole, with $r_+\gg \ell_p^2$, the near-horizon approximation used for the island configuration is therefore not
applicable to the islandless case. Substituting the appropriate matter entropy expression into the generalized entropy gives the same result as Eq.~\eqref{eq:withot island result}. Thus, the near-horizon placement of the cutoff surface does not alter the islandless result; the radiation entropy continues to grow with time in the absence of an island
\begin{equation}\label{eq:withot island result}
    S_{\text{gen}}\approx\frac{c}{12}\kappa t_A.
\end{equation}

\subsection{Page time and Scrambling time}\label{sec 5}

We now study the Page time and scrambling time for a non-extremal Kerr black hole. The Page time characterizes the transition between the no-island and island phases of the radiation entropy. For an evaporating black hole, this transition marks the onset of the decrease in the radiation entropy. For an eternal black hole coupled to an external bath, the radiation entropy instead
approaches a constant value after the island becomes dominant. Thus, the Page time can be estimated by comparing the no-island and island contributions to the generalized entropy. It is approximately determined by the intersection of the corresponding entropy curves. Comparing eqn \eqref{eq:sgen final} and eqn \eqref{eq:withot island result}, we obtain
\begin{equation}
    t_{\text{Page}}=\frac{12}{c\kappa}S_{BH}
\end{equation}
\begin{equation}\label{eq:Page time}
    =\frac{24\pi}{c G}\frac{(r_+^2+a^2)^2}{(r_+-r_-)}
    =\frac{12\pi}{cG}
    \frac{\Bigg[a^2+\left(M+\sqrt{M^2-a^2}\right)^2\Bigg]^2}
    {\sqrt{M^2-a^2}}.
\end{equation}
In the non-rotating limit, $a=0$, the Kerr geometry reduces to the Schwarzschild geometry, and our expression reduces accordingly. The result is also consistent with the previous analysis of the Kerr black hole reported in \cite{Wang:2024itz}.

\begin{figure}[h]
    \centering 
    \includegraphics[width=0.8\textwidth]{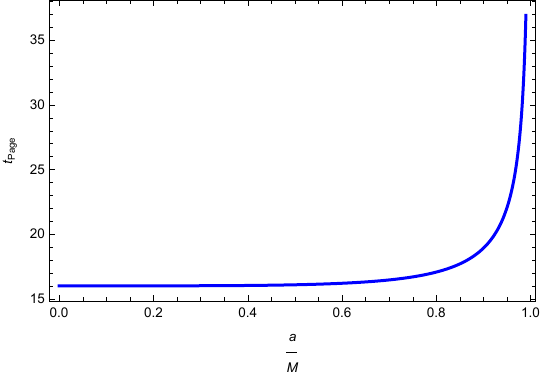}
    \caption{Page time as a function of the angular momentum
    parameter $a$, in units of $\frac{12\pi}{cG}$.
    It increases with rotation and diverges as the extremal
    limit is approached.}
    \label{fig:Page time vs a}
\end{figure}

As shown in Fig.~\ref{fig:Page time vs a}, the Page time increases with the angular momentum. This behaviour follows from the dependence of the surface gravity on the rotation parameter. In particular, as $a$ approaches its extremal value, the surface gravity tends to zero and the Page time
diverges. The extremal case requires a separate analysis, which we present in the subsequent section.\\
We next turn to the scrambling time. According to the Hayden--Preskill thought experiment \cite{Hayden:2007cs}, information that falls into a black hole can, in principle, be recovered from the outgoing Hawking radiation after a characteristic timescale known as the scrambling time. From the perspective of entanglement wedge reconstruction \cite{Penington:2019npb}, this timescale can be associated with the time required for the information to reach the boundary of the entanglement island. Consider an observer located at the cutoff surface $r=r_b$, who sends a light signal towards the black hole at time $t_1$. Assuming that the information carried by the signal can be decoded instantaneously once it enters the black hole, let $t_2$ denote the time at which it reaches the island at $r=r_a$. Along the relevant null direction, the relation between the two events is
\begin{equation}
    v_1(t_1,r_b)-v_2(t_2,r_a)
    =[t_1+r_{*}(r_b)]-[t_2+r_{*}(r_a)].
\end{equation}
The scrambling time is defined as the minimum time interval
$\Delta t=t_2-t_1$:
\begin{equation}
    t_{\text{scr}}=\text{min}[\Delta t]
    =\text{min}[(r_{*}(r_b)-r_{*}(r_a))
    -(v(t_1,r_b)-v(t_2,r_a))]
\end{equation}
\begin{equation}\label{eq:scrambling time}
   = r_{*}(r_b)-r_{*}(r_a).
\end{equation}
The scrambling time depends on the position of the island, which is itself sensitive to the location of the cutoff surface. We therefore consider two configurations: a cutoff surface away from the horizon and one located
close to the horizon.
\paragraph{Cutoff surface away from the horizon.}
We first consider a cutoff surface located away from the horizon. Substituting the island location given in Eq.~\eqref{eq:island position far from horizon} into the expression for the scrambling time \eqref{eq:scrambling time}, we obtain
\begin{equation}
    t_{scr}(a)=\frac{a^2+r_+^2}{(r_+-r_-)}
    \log\Bigg[\frac{24\pi r_+(r_A-r_+)}{cG}\Bigg]
\end{equation}
\begin{equation}
    \approx \frac{1}{2\kappa}\log S_{BH}
    +\frac{1}{2\kappa}
    \log\Bigg[\frac{24 r_+(r_A-r_+)}
    {c(r_+^2+a^2)}\Bigg]
\end{equation}
\begin{equation}\label{eq:scrambling time}
    \approx \frac{[\left(M+\sqrt{M^2-a^2}\right)^2+a^2]}
    {2\sqrt{M^2-a^2}}
    \log\Bigg[
    \frac{\pi[a^2+\left(M+\sqrt{M^2-a^2}\right)^2]}{G}
    \Bigg].
\end{equation}

\paragraph{Cutoff surface close to the horizon.}

We now place the cutoff surface close to the horizon. Using the island location given in
Eq.~\eqref{eq:island loc cut near}, we find
\begin{equation}
    t_{\text{scr}}(a)=\frac{a^2+r_+^2}{(r_+-r_-)}
    \log\Bigg[
    \frac{4 (r_A-r_+)^4(r_+^2+a^2)^2}
    {c^2G^2K^2r_+^2(r_+-r_-)^2}
    \Bigg]
\end{equation}
\begin{equation}
    \approx \frac{1}{\kappa}\log S_{BH}
    +\frac{1}{\kappa}
    \log\Bigg[
    \frac{2 (r_A-r_+)^2}
    {c\pi Kr_+(r_+-r_-)}
    \Bigg]
\end{equation}
\begin{equation}
    \approx \frac{\left[\left(M+\sqrt{M^2-a^2}\right)^2+a^2\right]}
    {\sqrt{M^2-a^2}}
    \log\Bigg[
    \frac{\pi[a^2+\left(M+\sqrt{M^2-a^2}\right)^2]}{G}
    \Bigg].
\end{equation}

\begin{figure}[h]
    \centering
    \begin{subfigure}{0.45\textwidth}
        \includegraphics[width=\linewidth]{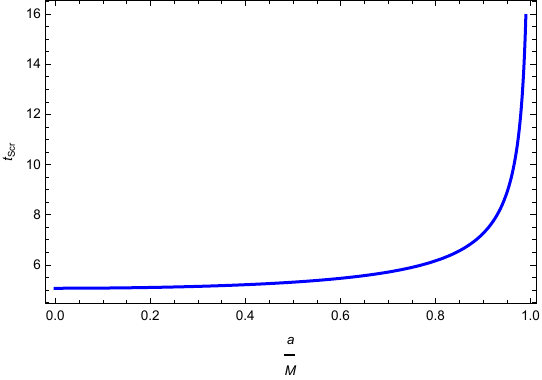}
        \caption{}
    \end{subfigure}
    \begin{subfigure}{0.45\textwidth}
        \includegraphics[width=\linewidth]{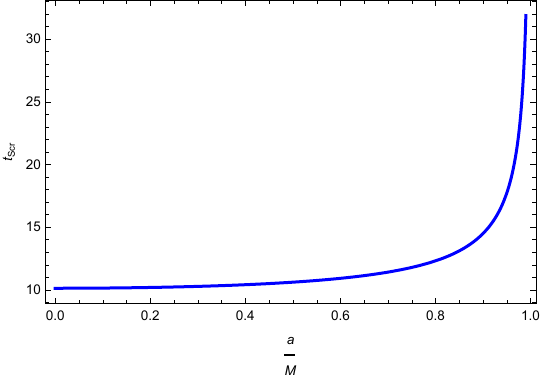}
        \caption{}
    \end{subfigure}
    \caption{Dependence of the scrambling time on the rotation parameter $a$, with $G=1$. The two panels show (a) the cutoff surface far from the horizon, (b) the cutoff surface close to the horizon.}
    \label{fig:plot of scrambling time}
\end{figure}

The results indicate that, within our approximations, the leading scrambling time is larger for a cutoff surface located close to the horizon. This behaviour depends not only on the cutoff position but also on the corresponding island location. Since the scrambling time is determined by the difference between the tortoise coordinates of these two locations, its dependence on the cutoff position is nontrivial. In particular, for a cutoff surface away from the horizon, we obtain the leading behaviour $t_{\mathrm{scr}}\simeq \frac{1}{2\kappa}\log S_{\mathrm{BH}}$, which exhibits the familiar logarithmic dependence on the black-hole entropy discussed in the Hayden--Preskill scenario \cite{Sekino:2008he,Wang:2022gbl}. In contrast, for a cutoff surface close to the horizon, the leading behaviour becomes $t_{\mathrm{scr}}\simeq \frac{1}{\kappa}\log S_{\mathrm{BH}}.$ Thus, the leading coefficients differ between the two configurations, reflecting the distinct island locations obtained in the two cutoff regimes.

In both configurations, the scrambling time has a logarithmic dependence on the black-hole entropy, whereas the Page time is proportional to the black-hole entropy. Consequently, for a large black hole and away from extremality, the scrambling time is parametrically smaller than the Page time. Similarly, for a special case $a=0$ , the above results also consistent with previous studies \cite{Wang:2024itz,Wang:2021woy}.

Finally, Fig.~\ref{fig:plot of scrambling time} displays the variation of the scrambling time with the angular momentum. Similar to the Page time, it increases as the rotation parameter (angular momentum) approaches its extremal value. The divergence in this limit signals the breakdown of the non-extremal approximation. We therefore examine the extremal case separately in the following section.

\section{Extremal Kerr black hole}\label{sec 6}
\begin{figure}[t]
    \centering
    \includegraphics[width=1.1\textwidth]{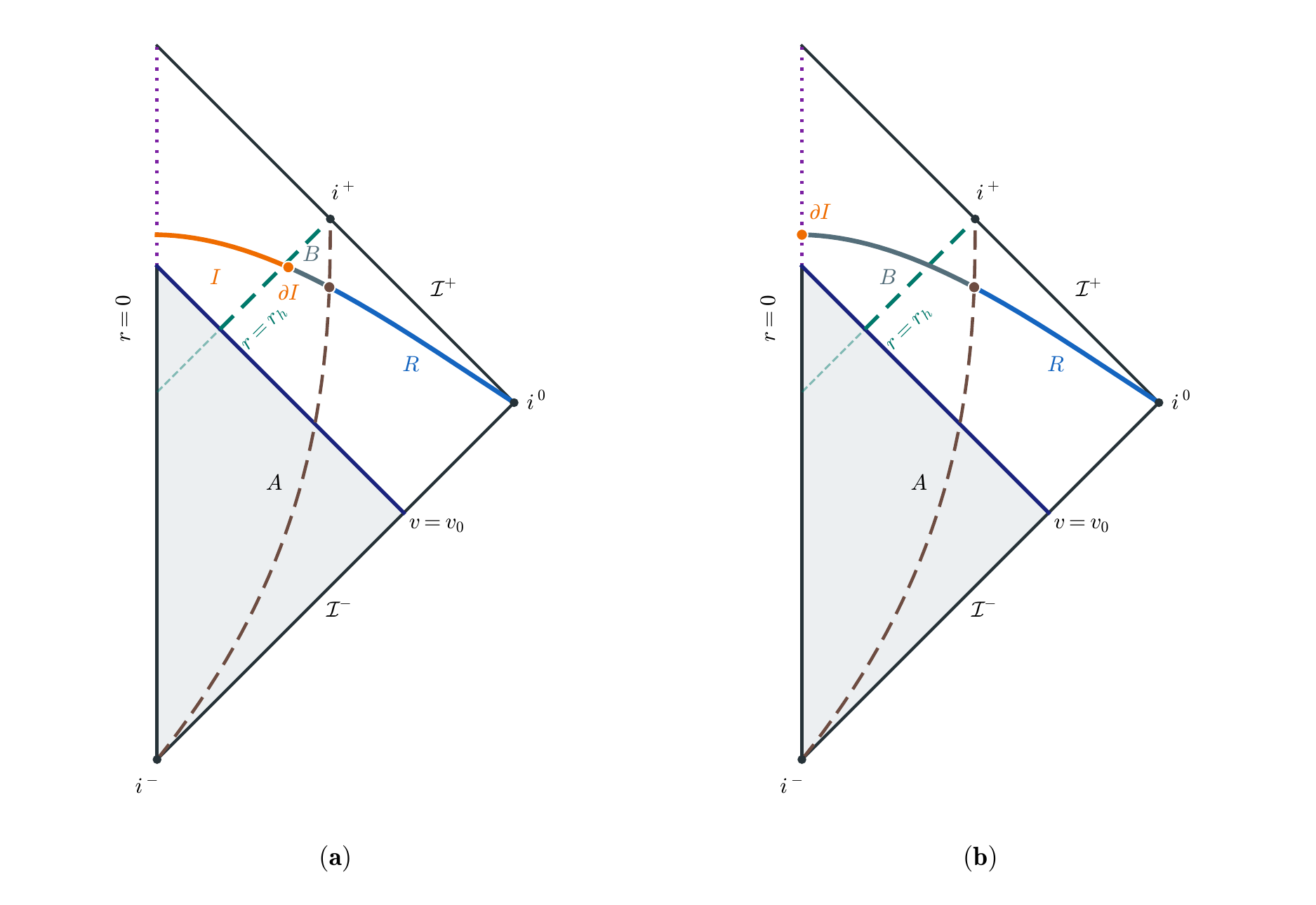}
 \caption{(a) Penrose diagram of an extremal Kerr black hole formed through
the collapse of a spherical null shell at $v=v_0$.
An observer at the cutoff surface (brown dashed curve) collects
Hawking radiation in the region $R$. The orange curve represents
the island, with $\partial I$ denoting its boundary.
(b) Configuration without an island, corresponding to the
no-island prescription, which can equivalently be obtained by
setting $r_I=0$.}
    \label{fig:Penrose_diagram_extremal}
\end{figure}

\subsection{Connecting condition}

We now turn to the extremal Kerr black hole and investigate the corresponding entanglement entropy. An important point is that the extremal case cannot, in general, be obtained simply by taking the extremal limit of the non-extremal
result. This is because extremal and non-extremal black holes have different global spacetime structures, as reflected in their respective Penrose diagrams. Therefore, the extremal case requires an independent treatment. For this purpose, we consider a one-sided extremal Kerr black hole formed through the collapse of a spherical null shell. Its Penrose diagram is presented in Fig.\eqref{fig:Penrose_diagram_extremal}. We work with the ``in'' vacuum state and use the corresponding
connecting condition to evaluate the generalized entropy, $S_{\text{gen}}$.\\
For an extremal Kerr black hole, the horizon radii coincide, with $r_+=r_-=M=a$. We denote the common horizon radius by $r_h$. Starting from the reduced two-dimensional metric given in Eq.~\eqref{eq: effective metric}, the tortoise coordinate $r^*(r)$ is defined by
\begin{equation}
    r^{*}=\int \frac{dr}{f(r)}=\int dr\frac{r^2+r_h^2}{(r-r_h)^2}=r-\frac{2r_h^2}{r-r_h}+2r_h\ln\Bigg|\frac{r-r_h}{r_h}\Bigg|.
\end{equation}
We introduce the double-null coordinates in the exterior and interior regions as
\begin{equation}
    u=t-r^*, v=t+r^* \ (\text{Exterior region}).
\end{equation}
\begin{equation}
    u_{\text{in}}=t-r_{\text{in}}, v=t+r_{\text{in}} \ (\text{Interior region}).
\end{equation}
Before the collapse, the geometry is described by the Minkowski metric. In double-null coordinates, the line element in this region takes the form
\begin{equation}
    ds^2=-du_{\text{in}}dv.
\end{equation}
After the collapse, the exterior geometry is described by the extremal Kerr metric. Its effective two-dimensional line element in double-null coordinates is
\begin{equation}
    ds^2=-f(r)dudv=-\frac{(r-r_h)^2}{r^2+r^2_h}dudv.
\end{equation}
Matching the coordinates across the collapsing null shell gives the following connecting condition for the extremal case:
\begin{equation}\label{conn con extr}
    u=u_{in}+\frac{8r_h^2}{v_0-u_{\text{in}}-2r_h}-4r_h\ln\Bigg|\frac{v_0-u_{in}-2r_h}{2r_h}\Bigg|.
\end{equation}
We next examine the late-time behaviour of this relation for the two possible locations of the island. For an island outside the horizon, we have
$u_{\text{in}}<v_0-2r_h$. In the late-time limit
$u\to\infty$, Eq.~\eqref{conn con extr} gives
\begin{equation}
    u_{\text{in}}\approx v_0-2r_h-\frac{8r^2_h}{u}.
\end{equation}
For an island inside the horizon, the corresponding condition is $u_{\text{in}}>v_0-2r_h$. Assuming that the island is located sufficiently close to the horizon, so that $u_{\text{in}}\approx v_0-2r_h$, the same leading-order relation is obtained, but with $u\to-\infty$:
\begin{equation}
    u_{\text{in}}\approx v_0-2r_h-\frac{8r^2_h}{u}.
\end{equation}
Using these late-time relations, we obtain the derivative of the coordinate transformation as
\begin{equation}
    \frac{du}{du_{\text{in}}}=\frac{u^2}{8r_h^2}.
\end{equation}
Consequently, the conformal factor transforms according to the coordinate relation between the ``in'' and ``out'' regions. It is given by
\begin{equation}
    e^{2\rho(u_{\text{in}},v)}=e^{2\rho(u,v)}\frac{du}{du_{\text{in}}}=\frac{(r-r_h)^2}{(r^2+r_h^2)}\frac{(t-r^{*})^2}{8r_h^2}.
\end{equation}

\subsection{Cutoff surface away from the horizon}
\subsubsection{With island}

We first consider an extremal Kerr black hole with the cutoff surface located away from the horizon. As in the non-extremal case, the generalized entropy depends on the position of the cutoff surface. Using the matter entropy
expression in Eq.~\eqref{eq:sent formula cut off far}, we obtain
\begin{equation}\label{S gen for extr}
\begin{aligned}
S_{\mathrm{gen}}
={}& \frac{\pi(r_I^2+a^2)}{G}
+\frac{c}{6}\Bigg[
\ln\left|t_I-r_I^*-t_A+r_A^*\right|+\ln\left|t_I+r_I^*-t_A-r_A^*\right|\\
&
+\ln|r_A-r_h|-\frac{1}{2}\ln(r_A^2+r_h^2)
+\ln|r_I-r_h|
-\frac{1}{2}\ln(r_I^2+r_h^2)
\Bigg].
\end{aligned}
\end{equation}
We determine the island's time coordinate by extremizing the generalized entropy with respect to $t_I$. The resulting condition is
\begin{equation}
\frac{\partial S_{\text{gen}}}{\partial t_I}
=\frac{c (t_A - t_I)}
{3 (t_I-r_I^*  - t_A + r_A^*)
(t_I+r_I^*  - t_A - r_A^*)}=0,
\end{equation}
which fixes the island's time to be
\begin{equation}\label{tI=tA}
    t_I=t_A.
\end{equation}
Next, since the island boundary is expected to lie close to the extremal horizon, we parametrize its radial position as
\begin{equation}\label{rI approx}
    r_I=r_h(1+\alpha),
\end{equation}
where $|\alpha|\ll1$. Using Eq.~\eqref{tI=tA} and this near-horizon parametrization, the generalized entropy takes the form
\begin{equation}\label{S gen simplified}
\begin{aligned}
S_{\mathrm{gen}}
={}& \frac{\pi(r_I^2+a^2)}{G}
+\frac{c}{6}\Bigg[
2\ln\left|r_I^*-r_A^*\right|
+\ln|r_A-r_h|-\frac{1}{2}\ln(r_A^2+r_h^2)\\
&
+\ln|r_I-r_h|
-\frac{1}{2}\ln(r_I^2+r_h^2)
\Bigg].
\end{aligned}
\end{equation}
The radial position of the island is then obtained by extremizing Eq.~\eqref{S gen simplified} with respect to $r_I$. Expressing the result in terms of $\alpha$, we find
\begin{equation}
\begin{aligned}
\frac{\partial S_{\text{gen}}}{\partial r_I}
={}&\frac{2\pi r_h(1+\alpha)}{G}
+\frac{c[(1+\alpha)^2+1]}
{3\alpha^2\left[
r_h(1+\alpha)+2r_h\ln|\alpha|
-\frac{2r_h}{\alpha}-r_A^{*}
\right]}
\\
&+\frac{c}{6r_h}
\left[
\frac{(2+\alpha)}
{\alpha(2+\alpha(2+\alpha))}
\right]=0.
\end{aligned}
\end{equation}
Keeping terms only up to first order in $\alpha$, the extremality condition reduces to
\begin{equation}
    48\pi r_h^3\alpha-4cGr_h(1+4\alpha)
    +2cG(r_h+\alpha r_A^{*})=0.
\end{equation}
Solving for $\alpha$, we obtain
\begin{equation}
    \alpha=\frac{cG r_h}
    {24\pi r_h^3-8cGr_h+cGr_A^{*}}
    \approx
    \frac{cG r_h}{24\pi r_h^3+cGr_A^{*}}
    \ll1,
\end{equation}
where we have used $cG\ll r_h^2$ and $r_A\gg r_h$. The positive value of $\alpha$ indicates that the island boundary is located slightly outside the extremal horizon. Finally, substituting the extremal island position into
the generalized entropy, we obtain
\begin{equation}
\begin{aligned}
S_{\mathrm{gen}}\simeq{}&
\frac{2\pi r_h^2}{G}
+\frac{c}{6}\Bigg\{
2\ln\Bigg|
r_h\left(1+\frac{cG}{24\pi r_h^2}\right)
+2r_h\ln\left|\frac{cG}{24\pi r_h^2}\right|
-\frac{48\pi r_h^3}{cG}-r_A^*
\Bigg|+\ln\left|r_A-r_h\right|\\
&
-\frac{1}{2}\ln\Big|\left(r_A^2+r_h^2\right)\Big|+\ln\left|\frac{cG}{24\pi r_h^2}\right|
-\frac{1}{2}\ln\left[
\left(1+\frac{cG}{24\pi r_h^2}\right)^2+1
\right]
\Bigg\}.
\end{aligned}
\end{equation}
At leading order, the generalized entropy is therefore
\begin{equation}\label{eq:sgen final extremal}
    S_{\mathrm{gen}}
    \approx \frac{2\pi r_h^2}{G}
    =S_{\mathrm{BH}}.
\end{equation}

\subsubsection{Without island}

We now consider the configuration without an island, for which we set $r_I=0$. In this case, the gravitational contribution vanishes. Moreover, the conformal factor  satisfies $e^{2\rho(u_{\text{in}},v)(I)}=1$. The generalized entropy is therefore determined by the matter contribution and takes the form
\begin{equation}
S_{\rm no-island}
=\frac{c}{12}\ln\left[
d(A,I)^4 e^{2\rho_A}e^{2\rho_I}
\right].
\label{eq:kerr_no_island_mother}
\end{equation}
Using the corresponding coordinate transformations, we obtain
\begin{equation}
\begin{aligned}
    S_{\rm no-island}
    =\frac{c}{12}\ln\Bigg|
    &\left(v_0-2r_h-\frac{8r_h^2}{t_A-r_A^{*}}\right)^2
    (v_I-v_A)^2
    \frac{f(r_A)}{8r_h^2}
    (t_A-r_A^{*})^2
    \Bigg|.
\end{aligned}
\end{equation}
\begin{equation}
\begin{aligned}
   ={}&\frac{c}{6}\ln\Bigg|
    v_0-2r_h-\frac{8r_h^2}{t_A-r_A^{*}}
    \Bigg|
    +\frac{c}{6}\ln\Bigg|
    t_I-t_A-r_A^{*}
    \Bigg|
    +\frac{c}{12}\ln\Bigg|
    \frac{f(r_A)}{8r_h^2}
    \Bigg|
    +\frac{c}{6}\ln\left(t_A-r_A^{*}\right).
\end{aligned}
\end{equation}
In the late-time regime, $t_A\gg r_A\gg r_h$, the leading
behaviour of the radiation entropy becomes
\begin{equation}\label{eq:late time log growth}
    S_{\rm no-island}\approx \frac{c}{3}\ln t_{A}.
\end{equation}
Thus, in the absence of an island, the radiation entropy grows logarithmically with time, in contrast to the linear growth found in the non-extremal case [Eq. \eqref{eq:withot island result}]. This behaviour is interesting in light of previous studies of extremal black holes. In Ref.~\cite{Qu:2024sby}, the entanglement entropy and Page time for a one-sided extremal Reissner--Nordström black hole were argued to be ill-defined. In contrast, Ref.~\cite{Yu:2021rfg} studied the island prescription and Page curve for a rotating BTZ black hole, taking superradiant effects into account when addressing the extremal limit. Similarly, Ref.~\cite{Wang:2024itz} showed that the entropy of an eternal Kerr black hole remains well-defined and time-dependent in the extremal case, even in the absence of an island. In our analysis of a one-sided extremal Kerr black hole formed by gravitational collapse, we likewise find a well-defined no-island entropy that grows logarithmically at late times. We can therefore estimate the Page time by equating the no-island entropy with the Bekenstein--Hawking entropy. Using Eqs.~\eqref{eq:sgen final extremal} and \eqref{eq:late time log growth}, we obtain
\begin{equation}
    t_{\mathrm{Page}}\sim t_0\exp\left(\frac{3S_{\mathrm{BH}}}{c}\right),
\end{equation}
where \(t_0\) is a reference time scale \footnote{The reference time \(t_0\) is introduced to make the argument of the logarithm dimensionless. Its value depends on the choice of time units and the additive constant in the late-time entropy; it does not affect the leading exponential dependence of the Page time on \(\frac{S_{\mathrm{BH}}}{c}\).}. Thus, the logarithmic growth of the no-island entropy implies an exponentially large Page time in the black hole entropy. This is consistent with the sharp increase of the non-extremal Page time as the extremal limit, \(\frac{a}{M}\to 1\), is approached, as illustrated in Fig.~\ref{fig:Page time vs a}.

\subsection{Cutoff surface close to the horizon}
\subsubsection{With island}

When the cutoff surface is located close to the horizon, we
evaluate the matter entanglement entropy using Eq.~\eqref{eq:entropy near formula}, which
takes the form
\begin{equation}
    S_{\text{ent}}=-K c \ \frac{A}{L^2}.
\end{equation}
Here, $L$ denotes the effective distance between the cutoff
surface $A$ and the island boundary $I$. Since both
surfaces are assumed to lie in the near-horizon region, this
distance can be approximated by
\begin{equation}
    L\approx\sqrt{d(A,I)^2 e^{\rho_A}e^{\rho_I}}.
\end{equation}
Using the coordinate transformations obtained above, we
can express this distance as
\begin{equation*}
    L=\sqrt{\Bigg|[u_{\text{in}}(I)-u_{\text{in}}(A)]
    (v_A-v_I)
    \frac{(r_I-r_h)(r_A-r_h)}
    {\sqrt{(r_A^2+r_h^2)(r_I^2+r_h^2)}}
    \frac{(t_A-r_A^*)(t_I-r_I^*)}{8r_h^2}\Bigg|}.
\end{equation*}
Equivalently, we have
\begin{equation*}
   L =\sqrt{\Bigg|[t_I-r_I^*-(t_A-r_A^*)]
    [t_A+r_A^*-(t_I+r_I^*)]
    \frac{(r_I-r_h)(r_A-r_h)}
    {\sqrt{(r_A^2+r_h^2)(r_I^2+r_h^2)}}\Bigg|}.
\end{equation*}
The generalized entropy is then given by
\begin{equation}
    S_{\text{gen}}=\frac{\pi(r_I^2+r_h^2)}{G}
    -Kc \ \frac{A}{L^2}.
\end{equation}
Substituting the expression for $L$, we obtain
\begin{equation}\label{S gen for extre near}
    S_{\text{gen}} =\frac{\pi(r_I^2+r_h^2)}{G}
    -\frac{4 K c \pi r_A^2
    \sqrt{(r_A^2+r_h^2)(r_I^2+r_h^2)}}
    {[t_I-r_I^*-(t_A-r_A^*)]
    [t_A+r_A^*-(t_I+r_I^*)]
    (r_I-r_h)(r_A-r_h)}.
\end{equation}

We first determine the island time by extremizing the
generalized entropy with respect to $t_I$. The resulting
extremality condition is
\begin{equation}
    \frac{\partial S_{\text{gen}}}{\partial t_I}
    =\frac{8 c K \pi r_A^2 (t_I-t_A)}
    {(r_A-r_h)(r_h-r_I)
    (r_A^*-r_I^*+t_A-t_I)^2
    (r_A^*-r_I^*-t_A+t_I)^2}=0.
\end{equation}
This condition gives
\begin{equation}\label{tI=tA near cutoff}
    t_I=t_A.
\end{equation}
Using this result in Eq.~\eqref{S gen for extre near},
the generalized entropy becomes
\begin{equation}\label{S gen after tI=tA near}
S_{\text{gen}}=\frac{\pi (r_h^2+r_I^2)}{G}
-\frac{4 c K \pi r_A^2
\sqrt{(r_A^2+r_h^2)(r_I^2+r_h^2)}}
{(r_A-r_h)(r_I-r_h)(r_A^*-r_I^*)^2}.
\end{equation}

We now introduce a near-horizon parametrization for the
radial positions of the island boundary and the cutoff
surface. Following the near-horizon setup of
Ref.~\cite{Tong:2023nvi}, we write
$r_I=r_h(1+\varepsilon)$ and
$r_A=r_h(1+\sigma)$, with
$\varepsilon<\sigma\ll1$. The corresponding tortoise
coordinates are
\begin{equation}
    r_I^*=r_h(1+\varepsilon)-\frac{2r_h}{\varepsilon}
    +2r_h\ln|\varepsilon|,\qquad
    r_A^*=r_h(1+\sigma)-\frac{2r_h}{\sigma}
    +2r_h\ln|\sigma|
\end{equation}
Substituting these expressions into
Eq.~\eqref{S gen after tI=tA near}, we find
\begin{equation}\label{S gen near approx}
\begin{aligned}
S_{\text{gen}}
={}&\frac{\pi r_h^2\left[(1+\varepsilon)^2+1\right]}{G}
\\[4pt]
&-\frac{
4cK\pi r_h^4(1+\sigma)^2
\sqrt{\left((1+\varepsilon)^2+1\right)
\left((1+\sigma)^2+1\right)}
}{
\varepsilon\sigma r_h^2
\left[
r_h(1+\sigma)-\frac{2r_h}{\sigma}
+2r_h\ln|\sigma|
-r_h(1+\varepsilon)
+\frac{2r_h}{\varepsilon}
-2r_h\ln|\varepsilon|
\right]^2
}.
\end{aligned}
\end{equation}
In the regime under consideration, this expression further
simplifies to
\begin{equation}
    S_{\text{gen}}\approx
    \frac{\pi r_h^2\left[(1+\varepsilon)^2+1\right]}{G}
    -
    \frac{\pi c K \varepsilon\sigma (1+\sigma)^2
    \sqrt{((1+\varepsilon)^2+1)((1+\sigma)^2+1)}}
    {\left[\sigma-\varepsilon+
    \varepsilon\sigma\ln\left(\frac{\sigma}{\varepsilon}\right)
    \right]^2}.
\end{equation}

The island position is determined by extremizing
Eq.~\eqref{S gen near approx} with respect to
$\varepsilon$. This gives
\begin{equation}\label{eq:deriv of sgen}
\begin{aligned}
\frac{\partial S_{\mathrm{gen}}}{\partial \varepsilon}
={}&\frac{2\pi r_h^2(1+\varepsilon)}{G}-\frac{
cK\pi\varepsilon\sigma(1+\varepsilon)(1+\sigma)^2
\left[1+(1+\sigma)^2\right]
}{
\sqrt{\left[1+(1+\varepsilon)^2\right]
\left[1+(1+\sigma)^2\right]}
\left[-\varepsilon+\sigma+
\varepsilon\sigma\ln\left(\frac{\sigma}{\varepsilon}\right)
\right]^2
}
\\
&-\frac{
cK\pi\sigma(1+\sigma)^2
\sqrt{\left[2+\varepsilon(2+\varepsilon)\right]
\left[2+\sigma(2+\sigma)\right]}
}{
\left[-\varepsilon+\sigma+
\varepsilon\sigma\ln\left(\frac{\sigma}{\varepsilon}\right)
\right]^3
}
\left[
\varepsilon+\sigma+2\varepsilon\sigma
-\varepsilon\sigma\ln\left(\frac{\sigma}{\varepsilon}\right)
\right]
=0.
\end{aligned}
\end{equation}

To obtain an analytical solution, we use
$0<\varepsilon<\sigma\ll1$ and
$\sigma\pm\varepsilon\gg
\left|\sigma\varepsilon\ln(\sigma/\varepsilon)\right|$.
Under these approximations, Eq.~\eqref{eq:deriv of sgen}
reduces to
\begin{equation}
\frac{2\pi r_h^2}{G}
-\frac{
2\pi Kc\,\sigma(1+\sigma)^2(\varepsilon+\sigma)
}{
(\sigma-\varepsilon)^3
}=0.
\label{eq:kerr_approx}
\end{equation}

Introducing the variable $x=\frac{\varepsilon}{\sigma}$,
with $0<x<1$, we can rewrite
Eq.~\eqref{eq:kerr_approx} as
\begin{equation}\label{x eqn}
    \frac{(1+x)}{(1-x)^3}
    =\frac{\sigma r_h^2}{G K c(1+\sigma)^2}.
\end{equation}

\begin{figure}[h]
    \centering
    \includegraphics[width=0.5\textwidth]{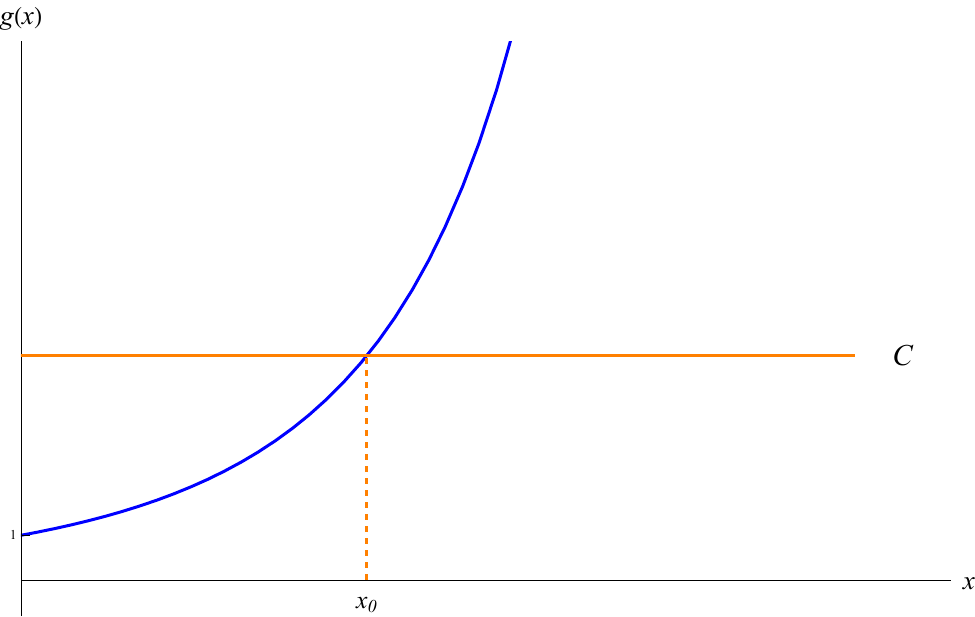}
    \caption{The plot of $g(x)$ as a function of $x$, where
$g(x)=\frac{1+x}{(1-x)^3}$ and
$C=\frac{\sigma r_h^2}{G K c(1+\sigma)^2}$.}
    \label{fig:gx_vs_x}
\end{figure}

The behaviour of the function on the left-hand side is
illustrated in Fig.~\ref{fig:gx_vs_x}.

The right-hand side of Eq.~\eqref{x eqn} is a constant.
Meanwhile, $g(x)=\frac{1+x}{(1-x)^3}$ is monotonically
increasing over the interval $0<x<1$, with its minimum
value $g(0)=1$ at the lower endpoint (see
Fig.~\ref{fig:gx_vs_x}). Consequently, a nontrivial island
solution exists only if the following condition is
satisfied:
\begin{equation}\label{constraint}
    C=\frac{\sigma r_h^2}{G K c(1+\sigma)^2}>1.
\end{equation}
When this condition holds, Eq.~\eqref{x eqn} admits a
unique solution $x=x_0$, corresponding to
$\varepsilon_0=x_0\sigma$. If the condition is not
satisfied, no nontrivial island solution is obtained
within the present approximation.

We have therefore established the existence of an island
solution in the near-horizon region
$\varepsilon<\sigma\ll1$, subject to the constraint
\eqref{constraint}, without requiring an explicit
expression for its position. We next evaluate the
corresponding radiation entropy under this condition.
From Eq.~\eqref{constraint}, we have
$\frac{\sigma r_h^2}{G}>Kc(1+\sigma)^2$.
Substituting $\varepsilon_0=x_0\sigma$ into
Eq.~\eqref{S gen near approx}, we obtain
\begin{equation}
    S_{\text{gen}}\approx
    \frac{\pi r_h^2\left[(1+\varepsilon_0)^2+1\right]}{G}
    -\frac{\pi c K \varepsilon_0\sigma^2
    \sqrt{((1+\varepsilon_0^2)+1)((1+\sigma^2)+1)}}
    {\left[\sigma-\varepsilon_0+
    \varepsilon\sigma\ln\left(\frac{\sigma}{\varepsilon_0}\right)
    \right]^2}.
\end{equation}
Using the constraint above, this gives
\begin{equation}
   S_{\text{gen}} > \frac{\pi r_h^2\left[(1+\varepsilon_0)^2+1\right]}{G}
    -\frac{\pi r_h^2\varepsilon_0\sigma^2
    \sqrt{((1+\varepsilon_0^2)+1)((1+\sigma^2)+1)}}
    {G(\sigma-\varepsilon_0)^2}.
\end{equation}
and hence
\begin{equation}
   S_{\text{gen}} =\frac{\pi r_h^2}{G}\left\{
    \left[(1+\varepsilon_0)^2+1)\right]
    -\frac{\sigma x_0
    \sqrt{((1+\varepsilon_0^2)+1)((1+\sigma^2)+1)}}
    {(1-x_0)^2}\right\}.
\end{equation}

It is useful to note that
$\frac{x_0}{(1-x_0)^2}<
\frac{1+x_0}{(1-x_0)^3}=C$.
This inequality allows us to establish the bound
\begin{equation}
    S_{\text{gen}}>
    \frac{\pi r_h^2}{G}
    \left\{2+2\varepsilon_0+\varepsilon_0^2-2\sigma C\right\}
    >\frac{2\pi r_h^2}{G}(1-\sigma C).
\end{equation}

On the other hand, the generalized entropy can also be
approximated as
\begin{equation}
    S_{\text{gen}}\approx
    \frac{\pi r_h^2\left[(1+\varepsilon)^2+1\right]}{G}
    -\frac{\pi c K \varepsilon\sigma(1+\sigma)^2
    \sqrt{((1+\varepsilon^2)+1)((1+\sigma^2)+1)}}
    {(\sigma-\varepsilon)^2}.
\end{equation}
which, upon using $\varepsilon_0=x_0\sigma$, becomes
\begin{equation}
   S_{\text{gen}} =\frac{2\pi r_h^2}{G}
    -\frac{2\pi c K x_0}{(1-x_0)^2}
    <\frac{2\pi r_h^2}{G}.
\end{equation}

These results show that the generalized entropy is close
to the Bekenstein--Hawking entropy. Thus, when the
constraint \eqref{constraint} is satisfied, the island
configuration provides an upper bound on the late-time
entropy of Hawking radiation. This bound is set by the
Bekenstein--Hawking entropy of the black hole, which
decreases during evaporation.

\subsubsection{Without island}

We now examine the case in which the cutoff surface is placed close to the horizon of the extremal Kerr black hole. In the no-island configuration, the island endpoint is located at $r_I=0$, which lies outside the near-horizon region (See the right panel of the Penrose diagram in Fig.~\ref{fig:Penrose_diagram_extremal}.). Therefore, the short-distance area-law prescription \eqref{eq:entropy near formula} cannot be used, and we must instead employ the general CFT expression for the entanglement entropy given in \eqref{eq:sent formula cut off far}. As a result, moving the cutoff surface closer to the horizon does not change the leading late-time behaviour of the no-island entropy, which continues to grow logarithmically, as found in Eq.~\eqref{eq:late time log growth}. Consequently, the Page time remains unchanged and is obtained by equating the no-island entropy with the black-hole entropy:
\begin{equation}
    t_{\mathrm{Page}} 
    \sim t_0\exp\left(\frac{3S_{\mathrm{BH}}}{c}\right).
    \label{eq:extremal_kerr_page_time}
\end{equation}

\section{Comparison with eternal Kerr black hole }\label{sec compareison}
In this section, we compare our results for a one-sided rotating Kerr black hole formed through gravitational collapse with previous studies of the island prescription for an eternal Kerr black hole \cite{Wang:2024itz,Yu:2021rfg}. We discuss the non-extremal and extremal cases separately. The main findings of our analysis are summarized in Table \ref{tab:one-sided-KR}. For a non-extremal Kerr black hole, when the cutoff surface is located far from the horizon, the island appears at a later time, with its boundary $\partial I$ lying inside and close to the horizon. In contrast, when the cutoff surface is placed near the horizon, the island boundary lies outside and close to it.\\

The extremal case exhibits a different behaviour. When the cutoff surface is far from the horizon, the island boundary $\partial I$ is located outside and close to the horizon, unlike the corresponding non-extremal configuration. When the cutoff surface is placed near the horizon, the island boundary again lies outside and close to the horizon. An important observation in our one-sided setup is that, even in the absence of an island, the generalized entropy grows logarithmically at late times, $S_{\mathrm{gen}}\simeq \frac{c}{3}\ln t_A.$ This behaviour contrasts with the result for the one-sided extremal Reissner--Nordström black hole, where the corresponding no-island entropy is ill-defined \cite{Qu:2024sby}.\\
For comparison, the results obtained for the eternal Kerr black hole in the Hartle--Hawking (HH) vacuum are summarized in Table \ref{tab:eternal-KR} \cite{Wang:2024itz}. This comparison highlights the differences between our analysis, which employs the ``in'' vacuum appropriate for a black hole formed through gravitational collapse, and previous studies based on the HH vacuum.
\begin{table}[htbp]
    \centering
    \renewcommand{\arraystretch}{1.7}
    \begin{tabular}{|c|c|c|c|}
        \hline
        Kerr black hole & without island & with island & Page time \\
        \hline
        non-extremal Kerr
        & $S_{\mathrm{gen}} \simeq \dfrac{c}{12}\kappa t_A$
        & $S_{\mathrm{gen}} \simeq S_{\mathrm{BH}}$
        & $t_{\mathrm{Page}} \simeq \dfrac{12}{c\kappa}S_{\mathrm{BH}}$ \\
        \hline
        extremal Kerr
        &$S_{\mathrm{gen}} \simeq \dfrac{c}{3}\ln t_A$
        & $S_{\mathrm{gen}} \simeq S_{\mathrm{BH}}$
        & $t_{\mathrm{Page}} \simeq t_0\exp(\frac{3 S_{\text{BH}}}{c})$ \\
        \hline
    \end{tabular}
    \caption{Summary of the radiation entanglement entropy and Page time for a one-sided Kerr black hole formed through gravitational collapse.}
    \label{tab:one-sided-KR}
\end{table}
\begin{table}[htbp]
    \centering
    \renewcommand{\arraystretch}{1.7}
    \begin{tabular}{|c|c|c|c|}
        \hline
        Kerr black hole & without island & with island & Page time \\
        \hline
        non-extremal Kerr
        & $S_{\mathrm{gen}} \simeq \dfrac{c}{3}\kappa t_A$
        & $S_{\mathrm{gen}} \simeq 2S_{\mathrm{BH}}$
        & $t_{\mathrm{Page}} \simeq \dfrac{6}{c\kappa}S_{\mathrm{BH}}$ \\
        \hline
        extremal Kerr
        & Linear with time $t_A$
        & $S_{\mathrm{gen}} \simeq S_{\mathrm{BH}}$
        & well-defined \\
        \hline
    \end{tabular}
    \caption{Summary of the entanglement entropy and Page time in the eternal Kerr black hole geometry.}
    \label{tab:eternal-KR}
\end{table}

\section{Conclusion and discussion}\label{sec 7}

In this work, we have investigated the black hole information problem in a four-dimensional one-sided rotating Kerr black hole. We have employed the ``in'' vacuum state, which is appropriate for describing one-sided dynamical black holes formed through gravitational collapse and is consistent with
the vacuum state in the initial Minkowski region. This distinguishes our analysis from previous studies based on the Hartle--Hawking state
\cite{Das:2026cjf,Wang:2024itz,Yu:2021rfg}, which is primarily suited to eternal black holes. Since the Kerr geometry is a non-spherically symmetric spacetime, we first showed that, in the near-horizon region, the quantum field
theory can be effectively described by a two-dimensional theory, as given in Eq.~\eqref{eq: effective metric}. This reduction allows us to describe the entanglement entropy within the framework of a two-dimensional conformal
field theory (CFT$_2$), thereby justifying the use of the s-wave approximation.\\
We have analysed non-extremal and extremal Kerr black holes separately, taking into account their distinct near-horizon geometries. For non-extremal Kerr black holes, we find that the island configuration depends on the location of the cutoff surface. When the cutoff surface is far from the horizon, the island boundary $\partial I$ lies inside and close to the horizon. In contrast, when the cutoff surface is placed near the horizon, $\partial I$ is located outside and close to it. In both configurations, the island contribution ensures that the radiation entropy remains bounded by the Bekenstein--Hawking entropy, reproducing the expected Page-curve behaviour. We have also obtained the Page time and scrambling time, given in
Eqs.~\eqref{eq:Page time} and \eqref{eq:scrambling time}, respectively. Their dependence on the angular momentum parameter is illustrated in Figs.~\ref{fig:Page time vs a} and \ref{fig:plot of scrambling time}. Both timescales increase as the rotation parameter increases and diverge in the extremal limit within the non-extremal description.\\
The extremal Kerr case exhibits some interesting differences. When the cutoff surface is far from the horizon, the island boundary $\partial I$ is located outside and close to the horizon, in contrast to the corresponding non-extremal
configuration. For a cutoff surface close to the horizon, an island solution exists only when certain constraints are satisfied, with its boundary again lying outside and near the horizon. We further find that, even in the absence
of an island, the radiation entropy in our effective two-dimensional description remains well-defined and grows logarithmically at late times. This behaviour differs from the extremal one-sided Reissner--Nordstr\"om case \cite{Qu:2024sby}, where the corresponding no-island entropy is ill-defined in the setup considered here.\\
Overall, our results demonstrate how rotation and the choice of cutoff surface influence the island location and the time evolution of radiation entropy in one-sided Kerr black holes. The analysis also highlights the differences between extremal and non-extremal geometries and their implications for the Page time and scrambling time. These findings provide further insight into the
island mechanism and the black hole information problem in dynamical rotating spacetimes. The comparison between our analysis of the one-sided Kerr black hole in the ``in'' vacuum state and previous studies of the eternal Kerr black hole in the Hartle--Hawking (HH) vacuum is presented in Section~\ref{sec compareison}.\\
There are several interesting directions for future work. One possibility is to extend the present analysis to Kerr-Newman black holes in order to investigate the combined effects of electric charge and angular momentum
on the island structure and Page curve. Another direction is to study the role of superradiance in the evolution of radiation entropy. Superradiant effects in eternal rotating BTZ black holes have previously been discussed in
\cite{Yu:2021rfg}. It would be interesting to investigate these effects in the present asymptotically flat setting, where the analysis does not require an external coupled thermal bath to absorb the Hawking radiation. Incorporating
superradiance into the island framework may provide further insight into how charge and rotation influence the Page curve and the black hole information problem.

\section*{Acknowledgments}

I would like to thank Harvendra Singh and Pabitra Tripathy for useful discussions and valuable comments on the manuscript. I acknowledge the Saha Institute of Nuclear Physics (SINP) and the Department of Atomic Energy (DAE), Government of India, for financial support through a doctoral fellowship. I also acknowledge the assistance of ChatGPT (OpenAI) in editing the language and improving the clarity of the manuscript. The image was also created with the assistance of Claude AI.
\appendix

\section{Cutoff surface close to the horizon with island boundary inside horizon}\label{app1}

We now consider the configuration in which the island boundary $\partial I$
lies close to and inside the event horizon. In this case, we use the interior Kruskal coordinate prescription given in Eq.~\eqref{eq:tor inside}. The effective distance between the two endpoints can then be approximated as
\begin{equation}
    L \approx
    \sqrt{d(A,I)^2 e^{\rho_A}e^{\rho_I}}.
\end{equation}
\begin{equation}
     =
    \frac{1}{\kappa}
    \sqrt{
    (U_A-U_I)(V_I-V_A)
    \sqrt{f(r_A)f(r_I)}
    e^{-\kappa(r_A^*+r_I^*)}
    }.
\end{equation}
The generalized entropy is given by
\begin{equation}
    S_{\mathrm{gen}}
    =
    \frac{\pi(r_I^2+a^2)}{G}
    -Kc\,\frac{A}{L^2}.
\end{equation}
Substituting the expression for $L$, we obtain
\begin{equation}
\begin{aligned}
    S_{\mathrm{gen}}
    ={}&
    \frac{\pi(r_I^2+a^2)}{G}-
    \frac{
    4\pi Kc\,r_A^2\kappa^2
    e^{\kappa(r_A^*+r_I^*)}
    }{
    (U_A-U_I)(V_I-V_A)
    \sqrt{f(r_A)f(r_I)}
    }.
\end{aligned}
\end{equation}

Substituting the explicit expression for the blackening factor from
Eq.~\ref{eq:tortoise co} and using the Kruskal coordinate transformations
given in Eq.~\ref{eq:Kruskal co trans}, the generalized entropy takes the
following form

\begin{equation}
\label{s gen case 2}
\begin{aligned}
S_{\mathrm{gen}}
={}&
\frac{\pi(r_I^2+a^2)}{G}
\\[4pt]
&-
\frac{
4\pi cK r_A^2\kappa^2
\sqrt{a^2+r_I^2}\sqrt{a^2+r_A^2}
}{
\sqrt{
(r_A-r_-)(r_I-r_-)
(r_A-r_+)(r_I-r_+)
}
}
\\[3pt]
&\quad\times
\frac{
e^{\kappa(r_A^*+r_I^*)}
}{
\left[
-e^{-\kappa(t_A-r_A^*)}
-e^{-\kappa(t_I-r_I^*)}
\right]
\left[
e^{\kappa(t_I+r_I^*)}
-e^{\kappa(t_A+r_A^*)}
\right]
}.
\end{aligned}
\end{equation}

After simplifying the time-dependent factors, we can rewrite this
expression as
\begin{equation}
\label{s gen case 2 simplified}
\begin{aligned}
S_{\mathrm{gen}}
={}&
\frac{\pi(r_I^2+a^2)}{G}
\\[3pt]
&-
\frac{
2\pi cK r_A^2\kappa^2
\sqrt{a^2+r_I^2}\sqrt{a^2+r_A^2}
}{
\sqrt{
(r_A-r_-)(r_I-r_-)
(r_A-r_+)(r_I-r_+)
}
}
\\[3pt]
&\quad\times
\frac{1}{
\sinh\!\left[\kappa(t_A-t_I)\right]
+
\sinh\!\left[\kappa(r_A^*-r_I^*)\right]
}.
\end{aligned}
\end{equation}

Since both the island boundary and the cutoff surface are located
close to the outer horizon, we introduce the parametrization
\begin{equation}
    r_A=r_+(1+\alpha),
    \qquad
    r_I=r_+(1-\beta),
    \qquad
    \alpha,\beta\ll1.
\end{equation}
Using the near-horizon expansion of the tortoise coordinate
given in Eq.~\eqref{eq:non-extre tor}, and retaining only the
leading-order contributions in $\alpha$ and $\beta$, we obtain
\begin{equation}
\label{eq:sgen for island inside}
\begin{aligned}
S_{\mathrm{gen}}
\simeq{}&
\frac{\pi(r_+^2+a^2)}{G}
-\frac{2\pi r_+^2}{G}\,\beta-
\frac{
\pi cK r_+(r_+-r_-)
}{
(r_+^2+a^2)
\left[
\alpha-\beta
+2\sqrt{\alpha\beta}\,
\sinh\!\left(\kappa(t_A-t_I)\right)
\right]
}.
\end{aligned}
\end{equation}

The quantum extremal surface is determined by extremizing the
generalized entropy with respect to the island's time coordinate $t_I$.
Differentiating Eq.~\eqref{eq:sgen for island inside}, we find
\begin{equation}
\begin{aligned}
\frac{\partial S_{\mathrm{gen}}}{\partial t_I}
={}&
-\frac{
2\pi cK(r_+-r_-)\kappa\sqrt{\alpha\beta}
}{
(r_+^2+a^2)
}
\frac{
\cosh\!\left[\kappa(t_A-t_I)\right]
}{
\left[
\alpha-\beta
+2\sqrt{\alpha\beta}\,
\sinh\!\left(\kappa(t_A-t_I)\right)
\right]^2
}
=0.
\end{aligned}
\end{equation}
The resulting extremality condition is
\begin{equation}
    \cosh\!\left[\kappa(t_A-t_I)\right]=0.
\end{equation}
Its solutions can be written as two branches:
\begin{equation}
    t_I
    =
    t_A-\frac{i\pi}{2\kappa}(4C_1+1),
    \qquad C_1\in\mathbb{Z},
\end{equation}
or
\begin{equation}
    t_I
    =
    t_A-\frac{i\pi}{2\kappa}(4C_1-1),
    \qquad C_1\in\mathbb{Z}.
\end{equation}
Thus, both branches yield complex values of the island time. Since there is no real solution to the time-extremality condition, we discard these saddles as unphysical within the present setup.\\
A similar conclusion is obtained in the extremal case: when the
cutoff surface is located close to the horizon, the time-extremality
condition does not admit a real solution for an island boundary
inside the horizon.

\bibliography{main.bib}
\bibliographystyle{JHEP}

\end{document}